\documentclass[aps,pra,floatfix,twocolumn,showpacs,superscriptaddress]{revtex4}%
\usepackage{amssymb}
\usepackage{amsmath}
\usepackage{amsfonts}
\usepackage{graphicx}
\def\Xint#1{\mathchoice
   {\XXint\displaystyle\textstyle{#1}}%
   {\XXint\textstyle\scriptstyle{#1}}%
   {\XXint\scriptstyle\scriptscriptstyle{#1}}%
   {\XXint\scriptscriptstyle\scriptscriptstyle{#1}}%
   \!\int}
\def\XXint#1#2#3{{\setbox0=\hbox{$#1{#2#3}{\int}$}
     \vcenter{\hbox{$#2#3$}}\kern-.5\wd0}}
\def\dashint{\Xint-}

\usepackage{hyperref}
\usepackage{epstopdf}

\begin{document}

\title{Bilayer superfluidity of fermionic polar molecules: many body effects}

\author{M.A. Baranov} 
\affiliation{Institute for Theoretical Physics, University of Innsbruck, A-6020 Innsbruck, Austria}
\affiliation{Institute for Quantum Optics and Quantum Information of the Austrian Academy of Sciences, A-6020 Innsbruck, Austria} 
\affiliation{RRC ``Kurchatov Institute'', Kurchatov Square 1, 123182 Moscow, Russia}
\author{A. Micheli}
\affiliation{Institute for Theoretical Physics, University of Innsbruck, A-6020 Innsbruck, Austria}
\affiliation{Institute for Quantum Optics and Quantum Information of the Austrian Academy of Sciences, A-6020 Innsbruck, Austria} 
\author{S. Ronen}
\affiliation{Institute for Theoretical Physics, University of Innsbruck, A-6020 Innsbruck, Austria}
\affiliation{Institute for Quantum Optics and Quantum Information of the Austrian Academy of Sciences, A-6020 Innsbruck, Austria} 
\author{P. Zoller}
\affiliation{Institute for Theoretical Physics, University of Innsbruck, A-6020 Innsbruck, Austria}
\affiliation{Institute for Quantum Optics and Quantum Information of the Austrian Academy of Sciences, A-6020 Innsbruck, Austria} 

\date{\today}

\begin{abstract}
We study the BCS superfluid transition in a single-component fermionic gas of dipolar particles
loaded in a tight bilayer trap, with the electric dipole moments polarized perpendicular to
the layers. Based on the detailed analysis of the interlayer scattering, we
calculate the critical temperature of the interlayer superfluid pairing
transition when the layer separation is both smaller (dilute regime) and of
the order or larger (dense regime) than the mean interparticle separation in
each layer. Our calculations go beyond the standard BCS approach and include
the many-body contributions resulting in the mass renormalization, as well as
additional contributions to the pairing interaction. We find that the
many-body effects have a pronounced effect on the critical temperature, and
can either decrease (in the very dilute limit) or increase (in the dense
and moderately dilute limits) the transition temperature as compared to the
BCS approach.
\end{abstract}

\pacs{67.85.-d, 03.75.Sc, 74.78.-w}

\maketitle

\section{Introduction}

Recent experiments have prepared quantum degenerate gases of homonuclear and
heteronuclear molecules in electronic and vibrational ground states \cite{Ni:08,Weidemueller:08,Danzl:10,NiOspelkausWang:10,Ospelkaus:10,Ni:10}.
Heteronuclear molecules, in particular, have large electric dipole moments
associated with the rotational excitations, The new feature of polar molecular
gases is thus the strong, anisotropic dipole-dipole interactions between the
molecules, which can be controlled with external electric 
fields \cite{Krems:09,Carr:09,Baranov:08,Lahaye:09}. When DC
electric fields are applied to polarize molecules, a major obstacle is given
by the increase of inelastic collision rates corresponding to chemical
reactions between the molecules, as in the case of KRb in the recent JILA
experiments. However, these can be strongly suppressed, and thus the gas
stabilized by tightly confining the molecules in a single plane of a quasi-2D
geometry\cite{Micheli}. This relies on the fact that for induced electric
dipole moments perpendicular to the plane of confinement, the dipolar forces
will be repulsive in-plane, thus suppressing short distance inelastic
collisions. Such a 2D confinement can be achieved by loading the gas of polar
molecules into a 1D optical lattice. This leads naturally to a setup of a
multilayer polar gas where, however, forces between dipoles in different
layers can be attractive, and the collapse is prevented by a sufficiently high
optical potential barrier. For a bilayer system this attraction can lead to
the formation of bound pairs of polar molecules, reminiscent of bilayer
excitons, and in a multilayer configuration to the formation of strings of
molecules. In particular, in a gas of single component fermions loaded in a
tight bilayer trap, as realized with KRb, this will give rise to an s-wave BCS
superfluid transition \cite{GoraSantos} (for $p$-wave pairing
in a monolayer of polar molecules see Refs. \cite{BruunTaylor} and \cite{Gorapwave}) . 
It is the purpose of this work to study this bilayer
BCS superfluid transition in some detail, in particular we go beyond Ref.~\cite{GoraSantos}
with emphasis on the inclusion of many body effects.

In the bilayer BCS superfluid the single species polar molecules in the two
layers provides the system with a two-component character, where two species
are particles on different layers coupled by the long-range dipole
interaction, allowing fermions from different layers interact in the $s$-wave
channel that is dominant at low energies. This interlayer interaction results
in very peculiar properties in both a two-body system: the existence of
interlayer bound states \cite{Simon}-\cite{Santos} and various regimes of the
interlayer scattering \cite{Santos}, and in a many-body system: interlayer BCS
pairing \cite{GoraSantos} and BCS-BEC crossover \cite{GoraSantos},
\cite{Zinner}. Based on a detailed analysis of various regimes of interlayer
and intralayer scattering, we extend the analysis of the interlayer superfluid
pairing beyond the standard BCS approach \cite{GoraSantos} by including
many-body contributions resulting in the mass renormalization, as well as in
additional contributions to the pairing interaction. We perform the
calculation of the critical temperature in the regime of a weak interlayer
coupling for the cases when the layer separation is both smaller (dilute
regime) and of the order or larger (dense regime) than the mean interparticle
separation in each layer. As found, the many-body effects have a pronounced
effect on the critical temperature, and could either decrease (in the very
dilute limit) or increases (in the dense and moderately dilute limits) the
transition temperature as compared to the BCS approach.

The paper is organized as follows. Sec.~II gives an overview of the problem 
and identifies the relevant parameters and parameter regimes. In Sec.~III we introduce 
the model describing bilayer pairing. Sec.~IV discusses two-particle bound states 
and scattering properties for the interlayer problem. 
 Sec.~V provides a theoretical treatment of many body effects in interlayer BCS pairing.
 Results for the critical temperature in the dilute and dense limit are summarized in Secs.~ VI and VII, respectively. 

\section{Overview of the problem and summary of the results}

The considered single-component fermionic bilayer dipolar system provides an
example of a relatively simple many-body system in which an entire range of
nontrivial many-body phenomena are solely tied to the dipole-dipole
interparticle interaction with its unique properties: long-range and
anisotropy. The long-range character provides an interparticle interaction in
single component Fermi gases inside each layer which otherwise would remain
essentially noninteracting. For the considered setup, this intralayer
interaction is always repulsive and gives rise to the crystalline phase for a
large density of particles. More important, the long-range dipole-dipole
interaction couples particles from different layers in a very specific form
resulting from the anisotropy of the interaction: two particle from different
layers attract each other at short, and repel each other at large  distances, respectively. The
potential well at short distances is strong enough to support at least one
bound state for any strength of interlayer coupling. For a weak coupling
between layers, the bound state is extremely shallow and has an exponentially
large size. However, in the intermediate and strong coupling cases the size of
the deepest bound state becomes comparable with the interlayer separation. In
the fermionic many-body system this behavior of the interlayer interaction
leads to a BCS state with interlayer Cooper pairs in the weak (interlayer)
coupling regime when the size of the bound state is larger than the
interparticle separation (in other words, the Fermi energy is larger than the
binding energy). With increasing interlayer coupling, this BCS state
smoothly transforms into a BEC state of tightly bound interlayer molecules
when the interparticle separation is larger than the size of the bound state.
Of course, the BEC regime and BEC-BCS crossover are only possible when the
mean interparticle separation in each layer is larger than the distance
between the layers.

In this paper we focus on the regime of a weak intra- and interlayer
interactions, that allows the usage of controllable calculations on the basis
of the perturbation theory, and consider in details the formation of the
interlayer BCS state. Before entering the technical derivation, it seems
worthwhile to briefly identify the relevant parameters and parameter regimes
for the bilayer many body system of single species fermionic dipoles. In
addition, we will point out, where the relevant results and discussions for
these parameter regimes can be found in later parts of the paper.

It follows from the previous discussion that the considered system is
characterized by three characteristic lengths: the dipolar length
$a_{d}=md^{2}/\hbar^{2}$, where $m$ is the mass of dipolar particles with the
(induced) dipole moment $d$, the interlayer separation $l$, and the mean
interparticle separation inside each layer $\sim k_{F}^{-1}$ with $k_{F}%
=\sqrt{4\pi n}$ being the Fermi wave vector for a 2D single-component
fermionic gas with the density $n$. Therefore, the physics of the system is
completely determined by two dimensionless parameters which are independent
ratios of the above lengths.

The first parameter $g$ is the ratio of the dipolar length and the interlayer
separation, $g=a_{d}/l$, and is a measure of the interlayer interaction
strength relevant for pairing. In experiments with polar molecules, the values
of the dipolar length $a_{d}$ is of the order of $10^{2}\div10^{4}%
\,\mathrm{nm}$: for a $^{40}\mathrm{K}^{87}\mathrm{Rb}$ with currently
available $d\approx0.3\,\mathrm{D}$ one has $a_{d}\approx170\,\mathrm{nm}$
(with $a_{d}\approx600\,\mathrm{nm}$ for the maximum value $d\approx
0.566\,\mathrm{D}$), and for $^{6}\mathrm{Li}^{133}\mathrm{Cs}$ with the
tunable dipole moment from $d=0.35\,\mathrm{D}$ to $d=1.3\,\mathrm{D}$ in an
external electric field $\sim1\,\mathrm{kV/cm}$ the value of $a_{d}$ varies
from $a_{d}\approx260\,\mathrm{nm}$ to $a_{d}\approx3500\,\mathrm{nm}$. For
the interlayer separation $l=500\,\mathrm{nm}$ these values of $a_{d}$
corresponds to $g\lesssim10$. 

The second parameter $k_{F}l$ measures the
interlayer separation in units of the mean interparticle distance in each
layer. This parameter can also be both smaller (dilute regime) and of the
order or larger (dense regime) than unity for densities $n=10^{6}\div
10^{9}\,\mathrm{cm}^{-2}$ (for, example, for $l=500\,\mathrm{nm}$ one has
$k_{F}l=1$ for $n\approx3\cdot10^{7}\,\mathrm{cm}^{-2}$).

The two parameters $g$ and $k_{F}l$ determine the regime of interlayer
scattering at typical energies of particles ($\sim$ Fermi energy
$\varepsilon_{F}=\hbar^{2}k_{F}^{2}/2m$), and their product, $gk_{F}l=a_{d}k_{F}$, 
being the ratio of the dipolar length and the mean
interparticle separation inside each layer, controls the perturbative
expansion in the system and, therefore, many-body effects. The existence of
different regimes of the interlayer scattering originates from two
characteristic features of the interlayer dipole-dipole interaction, as
discussed in the context of Eq.~(\ref{V2D}) below: the presence of the typical
range $\sim l$ beyond which the interaction is attactive, and of the
long-range dipole-dipole repulsive tail. As a result, the Fourier component of
the interaction (see Eq.~(\ref{V2DFourier}) below), decays exponentially for
large momentum $k\gg l^{-1}$, while it is proportional to $k$ for $k\ll
l^{-1}$, and, hence, vanishes for $k=0$. This leads to three different regimes
of scattering and, therefore, of the BCS pairing, depending on the relation
between $g$ and $k_{F}l$: regime a when $g<k_{F}l\lesssim1$, regime b when
$\exp(-1/g^{2})\ll k_{F}l<g<1$, and regime c when $\exp(-1/g^{2})\lesssim
k_{F}l\ll g<1$. The exponential factor in the last two formulae is related to
the size of an extremely shallow (in the limit of small $g$) interlayer bound
state (compare see Eq.~(\ref{Eb}) below).

\subsection{Regime a: $g<k_{F}l\lesssim1$} In this regime $g<k_{F}l\lesssim1$, and the scattering is
dominated by the first Born approximation (see Eq.~(\ref{Gamma_a}) and
(\ref{Gamma1s}) for the $s$-wave scattering amplitude for $k_{F}l\ll1$ and
$k_{F}l\lesssim1$, respectively). The critical temperature in this case will
be given below in Eqs.~(\ref{Tca}) and (\ref{Tcinterp}) for the dilute,
$k_{F}l\ll1$, and the dense, $k_{F}l\lesssim1$ cases, respectively. The ratio
of the critical temperature to the Fermi energy (chemical potential),
$T_{c}/\varepsilon_{F}$, can reach in this regime values of the order of
$0.1$, see Fig.~\ref{Fig_tau}, making an experimental realization of the
interlayer BCS pairing very promising. Note that the maximum of the ration
$T_{c}/\varepsilon_{F}$ corresponds to $k_{F}l\approx0.5$.

\subsection{Regime b: $\exp(-1/g^{2})\ll k_{F}l<g<1$} In the regime b, $\exp(-1/g^{2})\ll k_{F}l<g<1$, the
interlayer scattering is dominated by the second order Born contribution, see
Eq.~(\ref{Gamma_b}), and the critical temperature will be given below in Eq.
(\ref{Tcb}).

\subsection{Regime c: $\exp(-1/g^{2})\lesssim k_{F}l\ll
g<1$} Finally, in the regime c, $\exp(-1/g^{2})\lesssim k_{F}l\ll
g<1$, we recover the universal behavior for a two-dimensional low-energy
scattering, see Eq.~(\ref{Gamma_c}), with the typical inverse-logarithmic
dependence of the $s$-wave scattering amplitude on the energy. The critical
temperature is provided by Eq.~(\ref{Tcc}) and coincide with the critical
temperature in a two component gas with a short-range interaction with the
same Fermi energy and the bound state energy. Note that the values of the
critical temperature in the regimes b and c are much smaller ($T_{c}%
/\mu\lesssim10^{-3}$) than in the regime a. This makes an experimental
realization of the interlayer pairing in these regimes very challenging.

\section{The Model}

\begin{figure}[ptb]
\begin{center}
\includegraphics[width=.95\columnwidth]{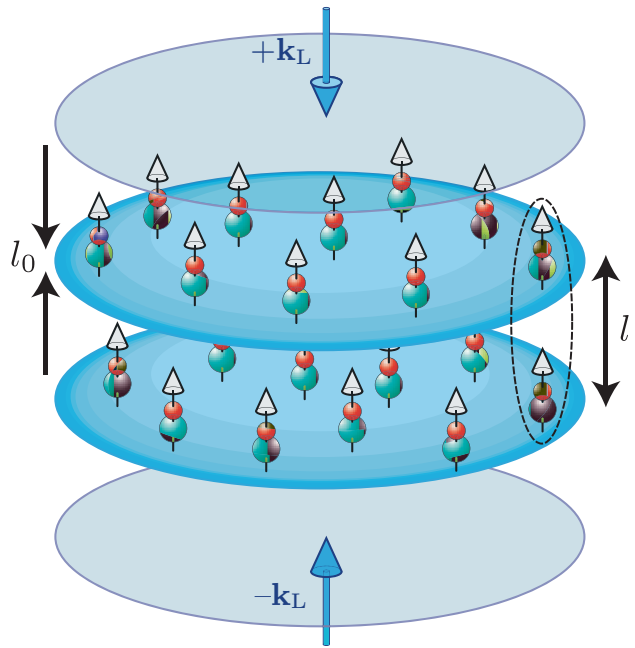}
\end{center}
\caption{The setup of the dipolar bilayer system: two layers with the thickness $l_0$ of a one-demensional 
optical lattice formed by two counterpropagating laser waves with wave vectors ${\bf k}_L$ and $-{\bf k}_L$, 
are filled with dipoles oriented perpendicular to the layers. The interlayer distance $l$ is $\pi /k_L$. 
An interlayer Cooper pair/molecule is schematically indicated by the dashed oval.}
\label{Fig_Setup}
\end{figure}

We consider a system of single-component polarized fermionic dipolar particles
(harmonically) confined in two infinite quasi-two-dimensional layers separated
by a distance $l$, which is much larger than the confinement length $l_{0}$ in
each layer, $l\gg l_{0}$. We assume that each layer has the same density $n$
of dipolar particles with mass $m$ and dipolar moment $d$ polarized along the
$z$-axis, which is perpendicular to the layers (see. Fig.~\ref{Fig_Setup}). The Hamiltonian of the system reads%
\begin{align}
&H    =\sum_{\alpha=\pm}\int d\mathbf{r}\hat{\psi}_{\alpha}^{\dagger
}(\mathbf{r})\left\{  -\frac{\hbar^{2}}{2m}\Delta+\frac
{1}{2}m\omega_{z}^{2}z_{\alpha}^{2}-\mu^{\prime}\right\}  \hat{\psi}_{\alpha
}(\mathbf{r})
 \nonumber\\
 & +\frac{1}{2}\sum_{\alpha,\beta}\int d\mathbf{r}d\mathbf{r}'%
\hat{\psi}_{\alpha}^{\dagger}(\mathbf{r})\hat{\psi}_{\beta
}^{\dagger}(\mathbf{r}')V(\mathbf{r}-\mathbf{r}'%
)\hat{\psi}_{\beta}(\mathbf{r}')\hat{\psi}_{\alpha}(\mathbf{r}), \label{H0}%
\end{align}
where $\alpha=\pm$ is the layer index, $z_\pm\equiv z\pm l/2$, $\hat{\psi}_{\alpha}(\mathbf{r})$ with $\mathbf{r}=(\boldsymbol{\rho},z)$ is the
field operator for fermionic dipolar particles ($\boldsymbol{\rho}=x\mathbf{e}%
_{x}+y\mathbf{e}_{y}$) on the corresponding layer $\alpha$, $\Delta=\Delta_{\boldsymbol{\rho}}+\partial^{2}/\partial_{z}^{2}$ is
the Laplace operator, $\omega_{z}$ is the confining frequency in each layer
such that $l_{0}=\sqrt{\hbar/m\omega_{z}}$,  and $\mu^{\prime}$ is the chemical
potential. The last term with
\[
V(\mathbf{r})=\frac{d^{2}}{r^{3}}\left(  1-3\frac{z^2}{r^2}\right)
\]
describes the intra- ($\alpha=\beta$) and interlayer ($\alpha\neq\beta$)
dipole-dipole interparticle interactions. Assuming a strong confinement,
$\hbar\omega_{z}\gg\mu',T$, where $T$ is the temperature, we can write%
\[
\hat{\psi}_{\alpha}(\mathbf{r})=\hat{\psi}_{\alpha}(\boldsymbol{\rho
})\phi_{0}(z_{\alpha})\equiv\hat{\psi}_{\alpha}(\boldsymbol{\rho})\frac{e^{-z_{\alpha}^{2}/2l_{0}^{2}}}%
{(\sqrt{\pi}l_{0})^{1/2}}
\]
and, therefore, reduce the Hamiltonian (\ref{H0}) to%
\begin{align}
&H_{2D}    =\sum_{\alpha=\pm}\int d\boldsymbol{\rho}\hat{\psi}_{\alpha}^{\dagger
}(\boldsymbol{\rho})\left\{  -\frac{\hbar^{2}}{2m}\Delta_{\boldsymbol\rho}-\mu\right\}  \hat{\psi}_{\alpha}(\boldsymbol{\rho})\label{H2D}\\
&  +\frac{1}{2}\sum_{\alpha,\beta}\int d\boldsymbol{\rho}d\boldsymbol{\rho}^{\prime
}\hat{\psi}_{\alpha}^{\dagger}(\boldsymbol{\rho})\hat{\psi}_{\beta}^{\dagger
}(\boldsymbol{\rho}^{\prime})V_{\alpha\beta}(\boldsymbol{\rho}-\boldsymbol{\rho}^{\prime
})\hat{\psi}_{\beta}(\boldsymbol{\rho}^{\prime})\hat{\psi}_{\alpha}(\boldsymbol{\rho
}),\nonumber
\end{align}
for a two-component fermionic field $\hat{\psi}_{\alpha}(\boldsymbol{\rho})$,
$\alpha=\pm$, with shifted chemical potential $\mu=\mu^{\prime}-\hbar
\omega_{z}/2$. 
The intracomponent (intralayer) interaction is 
\begin{align}
V_{\alpha\alpha}(\boldsymbol{\rho})&  =\int dz dz' V(\mathbf{r}-\mathbf{r}')\phi_{0}^{2}(z_{\alpha})\phi_{0}^{2}(z_{\alpha}')\nonumber\\
&  =\frac{d^{2}}{\sqrt{2\pi}l_{0}^{3}}\int_{0}^{\infty}d\xi\sqrt{\frac{\xi
}{(\xi+1)^{3}}}\exp\left(  -\xi\frac{\rho^{2}}{l_{0}^{2}}\right) \nonumber\\
&  =\frac{d^{2}}{\sqrt{8\pi}l_{0}^{3}}\exp\left(  -\frac{\rho^{2}}{4l_{0}^{2}%
}\right) \nonumber\\
&  \quad\left[  \left(  2+\frac{\rho^{2}}{l_{0}^{2}}\right)  \mathrm{K}%
_{0}\left(\frac{\rho
^{2}}{4l_{0}^{2}}\right)-\frac{\rho^{2}}{l_{0}^{2}}\mathrm{K}_{1}\left(\frac{\rho
^{2}}{4l_{0}^{2}}\right)\right]  ,\label{Vintra}
\end{align}
where $\mathrm{K}_{n}(z)$ is the modified Bessel Functions, and the
intercomponent (interlayer) one%
\begin{equation}
V_{+-}(\boldsymbol{\rho})=V_{-+}(\boldsymbol{\rho})\equiv V_{2D}(\boldsymbol{\rho})\approx
d^{2}\frac{\rho^{2}-2l^{2}}{(\rho^{2}+l^{2})^{5/2}}. \label{V2D}%
\end{equation}

The intralayer interaction $V_{++}$ is purely repulsive%
\[
V_{++}(\boldsymbol{\rho})\approx\left\{
\begin{array}
[c]{ccc}%
d^{2}/\rho^{3}&\mbox{for}&\rho\gg l_{0}\\
(\sqrt{2/\pi})(d^{2}/l_{0}^{3})\ln(l_{0}/\rho)&\mbox{for}&\rho\ll l_{0}%
\end{array}
\right.  .
\]
As a result, if the density $n$ is not too large (such that particles in each
layer are in a gas phase \cite{2Dcrystal1}, \cite{2Dcrystal2}), it leads only
to Fermi liquid renormalizations of the parameters of the Hamiltonian
(\ref{H2D})\ (effective mass, for example). The corresponding Fourier
transform reads%
\begin{equation}
\tilde{V}_{++}(\mathbf{k})=\sqrt{2\pi}\frac{4}{3}\frac{d^{2}}{l_{0}}+\tilde
{V}_{++}^{\prime}(\mathbf{k}),\label{V++}%
\end{equation}
where
\[
\tilde{V}_{++}^{\prime}(\mathbf{k})=-2\pi d^{2}k\exp(k^{2}l_{0}^{2}%
/2)[1-\operatorname{erf}(kl_{0}/\sqrt{2})]
\]
with $\operatorname{erf}(z)=(2/\sqrt{\pi})\int_{0}^{x}dse^{-s^{2}}$ being
the error function. In the considered limit $kl_{0}\ll1$, one simply has%
\begin{equation}
\tilde{V}_{++}^{\prime}(\mathbf{k})\approx-2\pi d^{2}k=-\frac{2\pi\hbar^{2}%
}{m}gkl.\label{V++approx}%
\end{equation}

\begin{figure}[ptb]
\begin{center}
\includegraphics[width=.95\columnwidth]{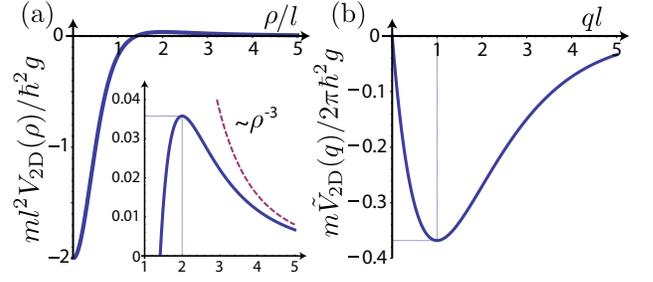}
\end{center}
\caption{The interlayer potential $V_{+-}(\rho)=V_{2D}(\rho)$ and its Fourier
transform $\tilde{V}_{2D}(q)$.}%
\label{Fig_Potential}%
\end{figure}

The effect of the interlayer interaction $V_{2D}(\boldsymbol{\rho})$ (see Fig.
\ref{Fig_Potential}) is more interesting. 
A peculiar property of $V_{2D}(\boldsymbol{\rho})$ is%
\begin{equation}
\int d\boldsymbol{\rho}V_{2D}(\boldsymbol{\rho})=0.\label{V2D0}%
\end{equation}
This means that its Fourier transform%
\begin{equation}
\tilde{V}_{2D}(\mathbf{q})=\int d\boldsymbol{\rho}V_{2D}(\boldsymbol{\rho}%
)e^{-i\mathbf{q }\boldsymbol{\rho}}=-\frac{2\pi\hbar^{2}}{m}gqle^{-ql}\label{V2DFourier}%
\end{equation}
vanishes for small $q$,%
\[
\tilde{V}_{2D}(\mathbf{q\rightarrow0})\approx-\frac{2\pi\hbar^{2}}%
{m}gql\rightarrow0.
\]
Hence, for sufficiently small energies, the interparticle scattering will be
dominated by higher order contributions in the Born expansion. Note also that
the linear dependence of $\tilde{V}_{2D}(\mathbf{q})$ on $q$ for $ql\ll1$ is
the consequence of the long-range power decay of $V_{2D}(\boldsymbol{\rho})$ for
large $\rho$ (the so-called anomalous contribution to scattering, see
\cite{LL3}). Another consequence of Eq.~(\ref{V2D0}) is that the interlayer
bound state, which always exists in the potential $V_{2D}(\boldsymbol{\rho})$
\cite{Simon}-\cite{Santos}, has extremely low binding energy for $g\ll1$.

From the point of view of many-body physics, however, the crucial observation
is that $\tilde{V}_{2D}(\mathbf{q})$ is negative for all $q$, signalling the
possibility of the interlayer pairing in the form of BCS pairs when the size
of the interlayer bound state $R_{b}$ is much large that the interparticle
separation, $R_{b}\gg n^{-1/2}$, or in the form of interlayer dimers for
$R_{b}<n^{-1/2}$ with some crossover in between (analogous to the BEC-BCS
crossover in 2D and 3D for two-component Fermi gases with short-range interactions).

The Hamiltonian (\ref{H2D}) is characterized by two parameters: $g=md^{2}%
/\hbar^{2}l,$which is the ratio of the dipolar length $a_{d}=md^{2}/\hbar^{2}$
and the interlayer spacing $l$, and $k_{F}l,$where $k_{F}=(2\pi n)^{1/2}$ is
the Fermi wave vector ($p_{F}=\hbar k_{F}$ is the Fermi momentum), which is
the ratio of the interlayer spacing $l$ and the average interparticle
separation in each layer. Note that the quantity $gk_{F}l=a_{d}k_{F}$ measures
the strength of the intralayer dipole-dipole interaction energy $d^{2}%
n^{3/2}\sim d^{2}k_{F}^{3}$ to the mean kinetic energy of particles
$\sim\varepsilon_{F}=\hbar^{2}k_{F}^{2}/2m$. In this paper we consider weakly
interacting gas of dipolar particles with $gk_{F}l<1$ in the dilute $k_{F}l<1$
and dense $k_{F}l\gtrsim1$ regimes.

\section{The Interlayer two-body problem: bound states and scattering
properties}

Let us first discuss the bound state and the scattering of two particles
interacting with the potential $V_{2D}(\rho)$ - interlayer two-body problem
(this problem was also addressed in Ref. \cite{Santos}, see, however,
discussion below). For this purpose we have to solve the 2D Sch\"{o}dinger
equation for the relative motion wave function $\psi(\boldsymbol{\rho})$%
\begin{equation}
\left\{  -\frac{\hbar^{2}}{2m_{r}}\Delta_{\boldsymbol\rho}+V_{2D}(\boldsymbol\rho)\right\}
\psi(\boldsymbol{\rho})=E\psi(\boldsymbol{\rho}), \label{Schroedinger0}%
\end{equation}
where $m_{r}=m/2$ is the reduced mass and the function $\psi(\boldsymbol{\rho})$
is regular for $\boldsymbol{\rho}\rightarrow0$. For the bound state solution with
$E=-E_{b}<0$, the wave function $\psi(\boldsymbol{\rho})$ should decay
exponentially for large $\rho$, while for the scattering wave function
$\psi_{\bf k}^{(+)}(\boldsymbol{\rho})$ with $E=\hbar^2k^{2}/m>0$, the boundary condition for
large $\rho$ reads%
\[
\psi_{\mathbf{k}}^{(+)}(\boldsymbol{\rho})\approx\exp(i\mathbf{k}\boldsymbol{\rho})-\frac
{f_{k}(\varphi)}{\sqrt{-8\pi ik\rho}}\exp(ik\rho),
\]
where $f_{k}(\varphi)$ is the scattering amplitude and $\varphi$ is the
azimuthal angle, $\mathbf{k}\boldsymbol{\rho}=k\rho\cos(\varphi)$. (Our definition of the
scattering amplitude correspond to that of Ref. \cite{PeytrovShlyapnikov},
which is differs by a factor of $-\sqrt{8\pi k}$ from the definition of Ref.
\cite{LL3}.)

\subsection{Bound state}

Writing the wave function for the relative motion of two particles in the form%
\[
\psi(\boldsymbol{\rho})=\chi_{m_z}(\rho)\exp(im_{z}\varphi),
\]
where $m_{z}$ is the magnetic quantum number, we obtain the following equation
for the radial wave function $\chi_{m_z}(\rho)$ of the bound state with the
binding energy $E_{b}$:%
\begin{equation}
\left[  \frac{d^{2}}{d\rho^{2}}+\frac{1}{\rho}\frac{d}{d\rho}-\frac{m_{z}%
^{2}}{\rho^{2}}-\frac{E_{b}+V_{2D}(\rho)}{\hbar^{2}/m}\right]  \chi
_{m_z}(\rho)=0, \label{Schroedinger}%
\end{equation}
and the function $\chi_{m_z}(\rho)$ should be regular for $\rho\rightarrow0$ and
decays exponentially for large $r$.

It is sufficient for our purposes to consider the azimuthally symmetric case
$m_{z}=0$, for which we consider two limiting cases $g\gg1$ and $g\ll1$. In
the first case, the potential $gv(r)$ supports several ($\sim g^{1/2}$) bound
states, and the lowest bound state has the binding energy $\varepsilon
_{b}\approx2(1-\sqrt{3/g})$ and size $r_{b}\sim(12g)^{-1/4}$ ($E_{b}%
=(\hbar^{2}/ml^{2})2g(1-\sqrt{3/g})$ and $R_{b}\sim l(12g)^{-1/4}$ in normal
units). In the opposite limit $g\ll1$, there is only one shallow bound state
(the existence of this bound state was proven in Ref. \cite{Simon}) with the
binding energy (see details of the derivation in Appendix A)%
\begin{equation}
E_{b}\approx\frac{4\hbar^{2}}{ml^{2}}\exp\left[ -\frac{8}{g^{2}}+\frac
{128}{15g}-\frac{2521}{450}-2\gamma+{\cal O}(g)\right]  ,\label{Eb}%
\end{equation}
where $\gamma\approx0.5772$ is the Euler constant, and with the size%
\[
R_{b}=\sqrt{\hbar^2/mE_{b}}\sim l\exp(4/g^2)\gg l.
\]
Note that the expression (\ref{Eb}) for the binding energy coincides with the
corresponding the one given in Ref. \cite{Santos} only to the leading order
($\sim1/g^{2}$). This is because the next order terms in the exponent ($\sim
g^{-1}$ and $\sim g^{0}$) are determined by the terms of the third and fourth
order in $g$ in the scattering amplitude (or in the wave function),
respectively (see Appendix A). ln Ref. \cite{Santos} however only terms up to
second order were taken into account and, therefore, only the leading term is
correct, see Figs. \ref{Fig_Boundstate1} and \ref{Fig_Boundstate2}.

\subsection{Scattering}

For the analysis of scattering it is convenient to introduce the vertex
function $\Gamma(E,\mathbf{k},\mathbf{k}^{\prime})$, where the arguments $E$,
$\mathbf{k}$,and $\mathbf{k}^{\prime}$ are independent of each other. This
function satisfies the following integral equation \cite{Taylor}
\begin{align}
\Gamma(E,\mathbf{k},\mathbf{k}^{\prime})  &  =V_{2D}(\mathbf{k}-\mathbf{k}%
^{\prime})+\int\frac{d\mathbf{q}}{(2\pi)^{2}}V_{2D}(\mathbf{k}-\mathbf{q}%
)\nonumber\\
&  \qquad\qquad\frac{1}{E-\hbar^{2}q^{2}/m+i0}\Gamma(E,\mathbf{q}%
,\mathbf{k}^{\prime}). \label{ScatteringIntegralEquation}%
\end{align}
The off-shell scattering amplitude%
\[
f_{\mathbf{k}}(\mathbf{k}^{\prime})=\frac{m}{\hbar^{2}}\int d\boldsymbol{\rho}%
\exp(-i\mathbf{k}^{\prime}\boldsymbol{\rho})V_{2D}(\boldsymbol\rho)\psi_{\mathbf{k}}%
^{(+)}(\boldsymbol{\rho})
\]
with $k\neq k^{\prime}$ can be obtained from $(m/\hbar^{2})\Gamma
(E,\mathbf{k},\mathbf{k}^{\prime})$ by putting $E=\hbar^{2}k^{2}/m$, and the
scattering amplitude $f_{k}(\varphi)$, where $\varphi$ is the angle between
$\mathbf{k}$ and $\mathbf{k}^{\prime}$, corresponds to $(m/\hbar^{2}%
)\Gamma(E,\mathbf{k},\mathbf{k}^{\prime})$ with $E=\hbar^{2}k^{2}/m=\hbar
^{2}k^{\prime2}/m$.

The iteration of Eq.~\eqref{ScatteringIntegralEquation} up to the fourth order
term in $\tilde V_{2D}({\bf q})$ reads%
\begin{widetext}%
\begin{align}\label{ScatteringForthOrderExpansion}
\Gamma(E,\mathbf{k},\mathbf{k}^{\prime})  &  =\tilde{V}_{2D}(\mathbf{k}%
-\mathbf{k}^{\prime})+\int\frac{d\mathbf{q}}{(2\pi)^{2}}\frac{\tilde{V}%
_{2D}(\mathbf{k}-\mathbf{q})\tilde{V}_{2D}(\mathbf{q}-\mathbf{k}^{\prime}%
)}{E-\hbar^{2}q^{2}/m+i0} +\int\frac{d\mathbf{q}_{1}d\mathbf{q}_{2}}%
{(2\pi)^{4}}\frac{\tilde{V}_{2D}(\mathbf{k}-\mathbf{q}_{1})\tilde{V}%
_{2D}(\mathbf{q}_{1}-\mathbf{q}_{2})\tilde{V}_{2D}(\mathbf{q}_{2}%
-\mathbf{k}^{\prime})}{(E-\hbar^{2}q_{1}^{2}/m+i0)(E-\hbar^{2}q_{2}^{2}%
/m+i0)}\nonumber\\
&  \qquad+\int\frac{d\mathbf{q}_{1}d\mathbf{q}_{2}d\mathbf{q}_{3}}{(2\pi)^{6}%
}\frac{\tilde{V}_{2D}(\mathbf{k}-\mathbf{q}_{1})\tilde{V}_{2D}(\mathbf{q}%
_{1}-\mathbf{q}_{2})\tilde{V}_{2D}(\mathbf{q}_{2}-\mathbf{q}_{3})\tilde
{V}_{2D}(\mathbf{q}_{3}-\mathbf{k}^{\prime})}{(E-\hbar^{2}q_{1}^{2}%
/m+i0)(E-\hbar^{2}q_{2}^{2}/m+i0)(E-\hbar^{2}q_{3}^{2}/m+i0)}+\ldots
\nonumber\\
&  \equiv\Gamma^{(1)}(E,\mathbf{k},\mathbf{k}^{\prime})+\Gamma^{(2)}%
(E,\mathbf{k},\mathbf{k}^{\prime})+\Gamma^{(3)}(E,\mathbf{k},\mathbf{k}%
^{\prime})+\Gamma^{(4)}(E,\mathbf{k},\mathbf{k}^{\prime})+\ldots.
\end{align}%
\end{widetext}

We now estimate the leading contributions of these terms in the small energy
limit $k\sim k^{\prime}\sim\sqrt{mE/\hbar^{2}}\ll1/l$:%
\begin{align}
\Gamma^{(1)}(E,\mathbf{k},\mathbf{k}^{\prime})  &  =\tilde{V}_{2D}%
(\mathbf{k}-\mathbf{k}^{\prime})\approx-\frac{2\pi\hbar^{2}}{m}g\left\vert
\mathbf{k}-\mathbf{k}^{\prime}\right\vert l,\label{Gamma1}\\
\Gamma^{(2)}(E,\mathbf{k},\mathbf{k}^{\prime})  &  \approx-\frac{2\pi\hbar
^{2}}{m}\frac{g^{2}}{4},\label{Gamma2}\\
\Gamma^{(3)}(E,\mathbf{k},\mathbf{k}^{\prime})  &  \approx-\frac{2\pi\hbar
^{2}}{m}\frac{4g^{3}}{15},\label{Gamma3}\\
\Gamma^{(4)}(E,\mathbf{k},\mathbf{k}^{\prime})  &  \approx-\frac{2\pi\hbar
^{2}}{m}\frac{g^{4}}{32}[\ln(\hbar^{2}/mEl^{2})+i\pi]. \label{Gamma4}%
\end{align}

The estimate for $\Gamma^{(1)}$ is trivial, the leading contributions to
$\Gamma^{(2)}$ and $\Gamma^{(3)}$ come from large $q$ ($q\gg k$) and large
$q_{1}$, $q_{2}$ ($q_{i}\gg k$) regions, respectively, and the leading
contributions to $\Gamma^{(4)}$ originates from large $q_{1}$ ($q_{1}\gg k$)
and $q_{3}$ ($q_{3}\gg k$) but small $q_{2}$ ($q_{2}\sim\sqrt{mE/\hbar^{2}}$).
Note that the next order terms (except those for $\Gamma^{(1)}$) have relative
magnitude of the order of $(kl)^{2}\ln(kl)$.

As already noted in Sec II. the above estimates show that there are three
different regimes of scattering for $g<1$ and $kl<1$ (dilute weakly
interacting regime for a many body fermionic system with $k\sim k_{F}$ and
$E\sim\varepsilon_{F}$): a. $g<kl<1$, b. $\exp(-1/g^{2})\ll kl<g<1$, and c.
$\exp(-1/g^{2})\lesssim kl\ll g<1$.

\emph{Regime a:} The leading contribution to scattering is given by the first
Born term
\begin{equation}
\Gamma_{\mathrm{a}}(E,\mathbf{k},\mathbf{k}^{\prime})\approx\Gamma
^{(1)}(E,\mathbf{k},\mathbf{k}^{\prime})\approx-\frac{2\pi\hbar^{2}}%
{m}g\left\vert \mathbf{k}-\mathbf{k}^{\prime}\right\vert l
\label{Gamma_a}%
\end{equation} valid for $g<kl<1$.

\emph{Regime b:} The scattering in this case is dominated by the second order
Born contribution%
\begin{equation}
\Gamma_{\mathrm{b}}(E,\mathbf{k},\mathbf{k}^{\prime})\approx\Gamma
^{(2)}(E,\mathbf{k},\mathbf{k}^{\prime})\approx-\frac{2\pi\hbar^{2}}{m}%
\frac{g^{2}}{4}\label{Gamma_b}%
\end{equation} valid for $\exp(-1/g^{2})\ll kl<g<1$.
In this case the scattering amplitude is momentum and energy independent and,
hence, is equivalent to a pseudopotential $V_{0}(\boldsymbol{\rho})=-(2\pi
\hbar^{2}/m)(g^{2}/4)\delta(\boldsymbol{\rho})$.

\emph{Regime c:} In this case the second and the fourth order contributions
become of the same order and one has to sum leading contributions from the
entire Born series. The result of this summation is%
\begin{equation}
\Gamma_{\mathrm{c}}(E,\mathbf{k},\mathbf{k}^{\prime})\approx\frac{2\pi
\hbar^{2}}{m}\frac{2}{\ln(E_{b}/E)+i\pi} \label{Gamma_c}%
\end{equation} valid for $exp(-1/g^{2})\lesssim kl\ll
g<1$,
where $E_{b}$ is the energy of the bound state from Eq. (\ref{Eb}). This
expression recovers the standard energy dependence of the 2D low-energy
scattering. The scattering amplitude $\Gamma_{c}(E,\mathbf{k},\mathbf{k}%
^{\prime})$ has a pole at $E=-E_{b}$, as it should be, and the real part of
$\Gamma_{c}(E,\mathbf{k},\mathbf{k}^{\prime})$, being zero at $E=E_{b}$,
changes from negative to positive values for $E>E_{b}$ and $E<E_{b}$,
respectively. Note that within the lowest order terms, one can write a unique
expression for the scattering amplitude for the three regimes in the form (for
more details see Appendix A)%
\[
\Gamma(E,\mathbf{k},\mathbf{k}^{\prime})\approx-\frac{2\pi\hbar^{2}}{m}\left[
g\left\vert \mathbf{k}-\mathbf{k}^{\prime}\right\vert l-\frac{2}{\ln
(E_{b}/E)+i\pi}\right]  .
\]

For later discussion we note that the scattering amplitude has both real and
imaginary parts. The relation between them can be established on the basis of
Eq. (\ref{ScatteringIntegralEquation}) by considering the imaginary part of
both sides of this equation,%
\begin{equation}
\operatorname{Im}\Gamma(E,\mathbf{k},\mathbf{k}^{\prime})=-\frac{m}{4\hbar
^{2}}\int\frac{d\varphi_{\mathbf{q}}}{2\pi}\Gamma^{\ast}(E,\mathbf{k}%
,\mathbf{q}_{E})\Gamma(E,\mathbf{q}_{E},\mathbf{k}^{\prime}), \label{ImGamma}%
\end{equation}
where $\left\vert \mathbf{q}_{E}\right\vert =\hbar^{-1}\sqrt{mE}$, the
integration is performed over the direction of this vector, and the complex
conjugate amplitude $\Gamma^{\ast}(E,\mathbf{k},\mathbf{k}^{\prime})$ obeys
Eq. (\ref{ScatteringIntegralEquation}) with $-i0$ in the denominator of the
integral term. This relation results in the unitarity condition for the
scattering matrix (optical theorem), and its validity in second order of the
perturbation theory is demonstrated in Appendix D. The analog of Eq.
(\ref{ImGamma}) for partial wave scattering amplitudes $\Gamma_{m}%
(E,k,k^{\prime})$ with azimuthal (magnetic) quantum number $m$ follows from
(\ref{ImGamma}) after integrating over the directions of $\mathbf{k}$ and
$\mathbf{k}^{\prime}$ with the proper angular harmonic. As an example, for the
$s$-wave scattering channel with%
\begin{equation}
\Gamma_{s}(E,k,k^{\prime})=\left\langle \Gamma(E,\mathbf{k},\mathbf{k}%
^{\prime})\right\rangle _{\varphi,\varphi^{\prime}} \label{def_Gamma_s}%
\end{equation}
we obtain%
\[
\operatorname{Im}\Gamma_{s}(E,k,k^{\prime})=-\frac{m}{4\hbar^{2}}\Gamma
_{s}^{\ast}(E,k,q_{E})\Gamma_{s}(E,q_{E},k^{\prime}).
\]
On the mass shell, $k=k^{\prime}=q_{E}=\hbar^{-1}\sqrt{mE}$, the above
relation reads%
\begin{equation}
\operatorname{Im}\Gamma_{s}(k)=-\frac{m}{4\hbar^{2}}\left\vert \Gamma
_{s}(k)\right\vert ^{2}. \label{ImGamma_s}%
\end{equation}
This implies that up to the second order one has%
\begin{equation}
\operatorname{Im}\Gamma_{s}(k)\approx-\frac{m}{4\hbar^{2}}[\operatorname{Re}%
\Gamma_{s}(k)]^{2}, \label{ImGamma_s_approx}%
\end{equation}
where
\begin{align}
\Gamma_{s}(k)  &  \approx\Gamma_{s}^{(1)}(k)=-\frac{2\pi\hbar^{2}}%
{m}gkl\left[  \mathbf{L}_{-1}(2kl)-\mathrm{I}_{1}(2kl)\right] \nonumber\\
&  \approx-\frac{2\pi\hbar^{2}}{m}gkl\frac{4}{\pi}\left(  1-\frac{\pi}%
{2}kl\right)  \label{Gamma1s}%
\end{align}
is just angular average of Eq. (\ref{Gamma1}). In Eq. (\ref{Gamma1s}),
$\mathbf{L}_{n}(z)$ and $\mathrm{I}_{n}(z)$ are the modified Struve and Bessel
functions, respectively, and the Taylor expansion in powers of $kl$ gives a
good approximation for $kl\lesssim0.2$.

\section{The Many-Body Problem}

Coming back to the many-body problem, we note that the amplitude of the
interlayer scattering in all three regimes is negative in the $s$-wave channel
(in the regime c this requires $E\sim\varepsilon_{F}\gg E_{b}$, which is
realistic in the limit $g\ll1$). This means that at sufficiently low
temperatures, the bilayer fermionic dipolar system undergoes a BCS pairing
transition into a superfluid state with interlayer $s$-wave Cooper pairs,
characterized by an order parameter $\Delta(\mathbf{p})\sim\left\langle
\hat{\psi}_{-}(\mathbf{p})\hat{\psi}_{+}(-\mathbf{p})\right\rangle $ with
$\hat{\psi}_{\alpha}(\mathbf{p})$ being the field operator in the momentum
space, which is independent of the azimuthal angle $\varphi$, $\Delta
(\mathbf{p})=\Delta(p)$.

\subsection{BCS approach to pairing}

The critical temperature $T_{c}$ of this transition is calculated from the
linearized gap equation. In the simplest BCS approach, which does not take
into account many-body effects (see below), this equation for the considered
system is (in what follows we will use the wave vector $\mathbf{k}$ instead of
the momentum $\mathbf{p}$)%
\begin{equation}
\Delta(\mathbf{k})=-\int\frac{d\mathbf{k}^{\prime}}{(2\pi)^{2}}\tilde{V}%
_{2D}(\mathbf{k}-\mathbf{k}^{\prime})\frac{\tanh(\xi_{k^{\prime}}/2T_{c}%
)}{2\xi_{k^{\prime}}}\Delta(\mathbf{k}^{\prime}), \label{BCSgapEq}%
\end{equation}
where $\xi_{k}=\hbar^{2}k^{2}/2m-\mu=\hbar^{2}(k^{2}-k_{F}^{2})/2m$ and
$\tilde{V}_{2D}(\mathbf{k}-\mathbf{k}^{\prime})$ is given explicitly by Eq.
(\ref{V2DFourier}). In the regime $k_{F}l\gtrsim1$, this equation can be
solved directly. For a dilute gas ($k_{F}l<1$) , however, the gap equation
(\ref{BCSgapEq}) with the bare interparticle interaction $\tilde{V}%
_{2D}(\mathbf{k})$ is not convenient because it mixes many-body physics (BCS
pairing) with the two-body one (scattering). In a dilute gas, they are
well-separated in momentum space: the pairing originates from the momenta of
the order of the Fermi momenta, $p\sim p_{F}$, while the two-particle
scattering is related to high momenta $p\sim\hbar/l\gg p_{F}$\thinspace\ that
correspond to short interparticle distances,\thinspace\ at which the presence
of other particles is irrelevant and physics is described by the two-particle
Schr\"{o}dinger equation. For the pairing problem, the two-body physics can be
taken into account by expressing the bare interparticle interaction $\tilde
{V}_{2D}(\mathbf{p}-\mathbf{p}^{\prime})$ in terms of the scattering amplitude
$\Gamma(E,\mathbf{k},\mathbf{k}^{\prime})$ using Eq.
(\ref{ScatteringIntegralEquation}). This results in the renormalized
(linearized) gap equation%
\begin{align}
\Delta(\mathbf{k})  &  =-\int\frac{d\mathbf{k}^{\prime}}{(2\pi\hbar)^{2}%
}\Gamma(2\mu,\mathbf{k},\mathbf{k}^{\prime})\left[  \frac{\tanh(\xi
_{k^{\prime}}/2T_{c})}{2\xi_{k^{\prime}}}\right. \nonumber\\
&  \left.  +\frac{1}{2\mu-\hbar^{2}k^{\prime2}/m+i0}\right]  \Delta
(\mathbf{k}^{\prime}), \label{renormalizedBCSgapEq}%
\end{align}
where we choose $E=2\mu$ for convenience. The contribution to the integral in
this equation comes only from momenta $p=\hbar k^{\prime}\sim p_{F}$ and,
therefore, the form (\ref{renormalizedBCSgapEq}) of the gap equation is more
suitable to describe the BCS pairing in a many-body dilute system.

In the regime of weak coupling characterized by a small parameter $\lambda
=\nu_{F}\Gamma\ll1$, where $\nu_{F}=m/2\pi\hbar^{2}$ is the density of state
on the Fermi surface ($k=k^{\prime}=k_{F}$), this equation can be solved by
expanding in powers of $\lambda$ (see below) or numerically. However, the
linearized BCS gap equation (\ref{renormalizedBCSgapEq}) can only be used for
the calculation of the leading contribution to the critical temperature,
corresponding to the terms $\sim\lambda^{-1}$ in the exponent. As was shown by
Gor'kov and Melik-Barkhudarov \cite{GM}, the terms of order unity in the
exponent affecting the preexponential factor in the expression for the
critical temperature, is determined by the next-to-leading order terms, which
depend on many-body effects. In the considered fermionic dipolar system, these
effects result in the appearance of the effective mass $m_{\ast}$ and the
effective interparticle interaction. The latter corresponds to the
interactions between particles in a many-body system through the polarization
of the medium - virtual creation of particle-hole pairs. The BCS pairing with
the account of the many-body effects can be viewed as a pairing of
quasiparticles of mass $m_{\ast}$ interacting with the effective interaction
$V_{\mathrm{eff}}$. Note that, in contrast to the Fermi gas with a short-range
interaction, in which the difference between the bare $m$ and the effective
$m_{\ast}$ masses (or, in other words, between particles and quasiparticles)
are of the second order in $\lambda$, in the dipolar system this difference is
typically of the first order in $\lambda$ due to the momentum dependence of
the dipolar interactions.

\subsection{The role of many-body effects}

Qualitatively the role of the many-body effects in the gap equation can be
understood as follows. After performing the integration over momenta, the gap
equation can be qualitatively written as%

\begin{equation}
1=\nu_{F}^{\ast}V_{\mathrm{eff}}\left[  \ln\frac{\mu}{T_{c}}+C\right]  ,
\label{eq:approx_eqTc}%
\end{equation}
where $\nu_{F}^{\ast}=m_{\ast}/2\pi\hbar^{2}$ is the density of states of
quasiparticles with the effective mass $m_{\ast}$, and we replace the
scattering amplitude $\Gamma$ with some effective interaction $V_{\mathrm{eff}%
}=\Gamma+\delta V$ with $\delta V$ being the many-body contribution to the
interparticle interaction. Note that the (large) logarithm $\ln\mu/T_{c}$
results from the integration over momenta near the Fermi surface, whereas the
momenta far from the Fermi surface contribute to the constant $C\sim1$. We can
now expand $\nu_{F}^{\ast}V_{\mathrm{eff}}$ in powers of $\lambda$ up to the
second order term, $\nu_{F}^{\ast}V_{\mathrm{eff}}=\lambda+a\lambda^{2}$,
where the first term results from the direct interparticle interaction and the
many-body effects (the difference between $m_{\ast}$ and $m$ together with
$\delta V$) contribute to the second term. In solving Eq.
(\ref{eq:approx_eqTc}) iteratively, we notice that $\lambda\ln\mu/T_{c}\sim1$
and, therefore, the terms $a\lambda^{2}\ln\mu/T_{c}$ and $\lambda C$ are of
the same order. As a result, both terms have to be taken into account for the
calculation of the critical temperature. It is easy to see that they
contribute to the preexponential factor in the expression for the critical
temperature. These contributions are usually called Gorkov-Melik-Barkhudarov
(GM) corrections \cite{GM}. It is important to notice that the many-body
effects appearing in Eq. (\ref{eq:approx_eqTc}) only in be combination with
the logarithm $\ln\mu/T_{c}$ that originates from momenta near the Fermi
surface. Therefore, it is sufficient for our purposes to consider the
many-body contributions only at the Fermi surface, and the renormalized
linearized gap equation with the account of the many-body effects reads
\begin{align}
\Delta(\mathbf{k})=  &  -\frac{m_{\ast}}{m}\int\frac{d\mathbf{k}^{\prime}
}{(2\pi)^{2}}\Gamma(2\mu,\mathbf{k},\mathbf{k}^{\prime})\left[  \frac
{\tanh(\xi_{k^{\prime}}/2T_{c})}{2\xi_{k^{\prime}}}\right. \nonumber\\
&  \qquad\left.  +\frac{1}{2\mu-\hbar^{2}k^{\prime2}/m+i0}\right]
\Delta(\mathbf{k}^{\prime}) \label{GapEq}\\
&  -\int_{k^{\prime}<\Lambda k_{F}}\frac{d\mathbf{k}^{\prime}}{(2\pi)^{2}%
}\delta V(\mathbf{k},\mathbf{k}^{\prime})\frac{\tanh(\xi_{k^{\prime}}/2T_{c}%
)}{2\xi_{k^{\prime}}}\Delta(\mathbf{k}^{\prime}), \nonumber%
\end{align}
where we introduce an upper cutoff $\Lambda k_{F}$ with $\Lambda\sim1$ for the
purpose of the convergency at large momenta. As discussed above, the exact
value of $\Lambda$ is not important because the large momenta contribution to
this integral has to be neglected.

\subsubsection{Effective mass}

\begin{figure}[ptb]
\begin{center}
\includegraphics[width=.9\columnwidth]{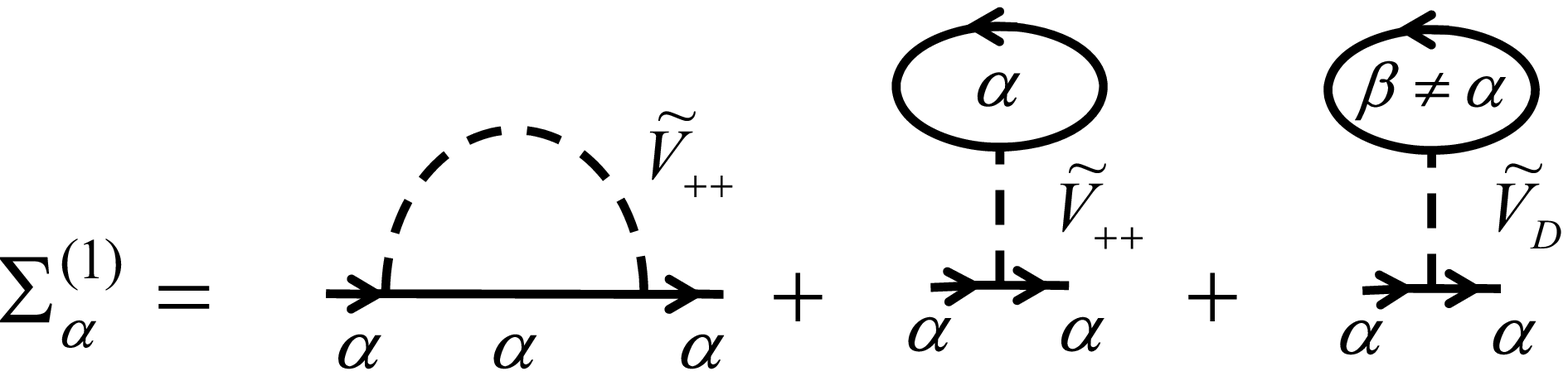}
\end{center}
\caption{The first order diagrams for the fermionic self-energy. Dashed lines
correspond to the dipole-dipole interactions. The first two diagrams contain
only the intralayer interaction, while the last one describes the effects of
the interlayer coupling.}%
\label{Fig_Sigma}%
\end{figure}

The contribution to the effective mass originates from the momentum and
frequency dependencies of the self-energy $\Sigma_{\alpha}(\omega,p)$ of
fermions (see, for example, Ref. \cite{LL9}), $m/m_{\ast}=(1+2m\partial
\Sigma/\partial p^{2})(1-\partial\Sigma/\partial\omega)^{-1}|_{p=p_{F}%
,\omega=0}$. In the considered case, the leading contributions to the
self-energy are shown in Fig. \ref{Fig_Sigma}, where the fermionic Green's function is%
\[
G_{\alpha}(\omega,p)=\frac{1}{\omega-\xi_{p}+i0~\mathrm{sign}(\xi_{p})}.
\]
It is easy to see that $\Sigma_{\alpha}^{(1)}(\omega,p)$ is frequency
independent, $\Sigma_{\alpha}^{(1)}(\omega,p)=\Sigma_{\alpha}(p)$, and the
momentum dependence results only from the first diagram that corresponds to
the exchange intralayer interaction (the momentum independent part of the
self-energy leads to unessential change of the chemical potential). The
analytical expression for $\Sigma_{\alpha}^{(1)}(p)$ reads ($p=\hbar k$)%
\begin{align}
\Sigma_{\alpha}^{(1)}(p) &  =-\int\frac{d\mathbf{q}}{(2\pi)^{2}}\left[
\tilde{V}_{++}(\mathbf{k}-\mathbf{q})-\tilde{V}_{++}(\mathbf{0})\right]
N(q)\nonumber\\
&\qquad+\tilde{V}_{2D}(0)n\nonumber\\
&  =-\int\frac{d\mathbf{q}}{(2\pi)^{2}}\left[  \tilde{V}_{++}(\mathbf{k}%
-\mathbf{q})-\tilde{V}_{++}(\mathbf{0})\right]  N(q),\nonumber
\end{align}
where $N(q)=\theta(k_{F}-q)$ is the Fermi-Dirac distribution for zero
temperature (the usage of $n_{q}$ at zero temperature is justified by the
exponential smallness of the critical temperature $T_{c}$) and $\tilde{V}%
_{+-}(\mathbf{k})$ is the Fourier transform of $\tilde{V}_{++}(\boldsymbol{\rho}%
)$, and straightforward calculations with the usage of Eqs. (\ref{V++}) and
(\ref{V++approx}) gives (see, for example, Ref. \cite{Das Sarma})%
\begin{equation}
\frac{m_{\ast}}{m}=1-\frac{4}{3\pi}a_{d}k_{F}=1-\frac{4}{3\pi}gk_{F}%
l.\label{effective_mass}%
\end{equation}

It is easy to see that higher order contribution will introduce small
parameters $gk_{F}l_{0}$ or $gk_{F}l$ and, therefore, can be
neglected.\newline

\subsubsection{Effective interparticle interaction}

\begin{figure}[phtb!]
\begin{center}
\includegraphics[width=.7\columnwidth]{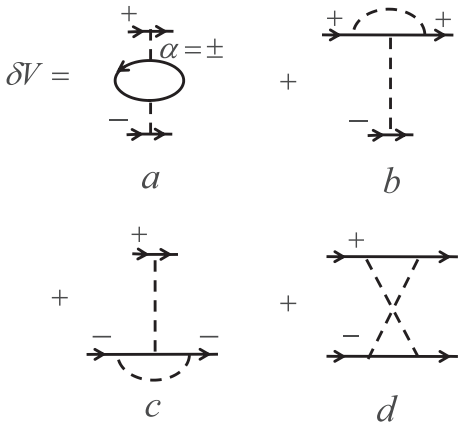}
\end{center}
\caption{The second-order contributions to the effective interlayer
interaction. Solid lines correspond to particles from different layers
(labeled by $+$ and $-$) and the dashed lines correspond to dipole-dipole
interactions. Note that the diagram d contains only the interlayer
interaction, while the diagrams a, b, and c have both the inter- and
intralayer interactions.}%
\label{Fig_deltaV}%
\end{figure}

Let us now discuss the many-body contributions to the effective interparticle
interaction. We consider first the case when the scattering of two particles
with energies of the order of the Fermi energy corresponds to the regime a
($g<k_{F}l<1$), and, hence, is well-controlled by the Born expansion in powers
of the bare interparticle interaction with the small parameter $\lambda
=\nu_{F}\Gamma\sim gk_{F}l$. As it was argued above, it is sufficient to
consider only the lowest (second order in the interparticle interactions)
many-body contributions to the effective interaction. These contributions are
shown in Fig. \ref{Fig_deltaV}, 
and the corresponding analytical expressions
read:
\begin{widetext}
\begin{align}
\delta V_{a}(\mathbf{k},\mathbf{k}^{\prime})  &  =2\int\frac{d\mathbf{q}%
}{(2\pi)^{2}}\frac{N(\mathbf{q}+\mathbf{k}_{-}/2)-N(\mathbf{q}-\mathbf{k}%
_{-}/2)}{\xi_{\mathbf{q}+\mathbf{k}_{-}/2}-\xi_{\mathbf{q}-\mathbf{k}_{-}/2}%
}\tilde{V}_{2D}(\mathbf{k}_{-})\tilde{V^{\prime}}_{++}(\mathbf{k}%
_{-}),\label{dVa}\\
\delta V_{b}(\mathbf{k},\mathbf{k}^{\prime})  &  =-\int\frac{d\mathbf{q}%
}{(2\pi)^{2}}\frac{N(\mathbf{q}+\mathbf{k}_{-}/2)-N(\mathbf{q}-\mathbf{k}%
_{-}/2)}{\xi_{\mathbf{q}+\mathbf{k}_{-}/2}-\xi_{\mathbf{q}-\mathbf{k}_{-}/2}%
}\tilde{V}_{2D}(\mathbf{k}_{-})\tilde{V^{\prime}}_{++}(\mathbf{q}%
-\mathbf{k}_{+}/2),\label{dVb}\\
\delta V_{c}(\mathbf{k},\mathbf{k}^{\prime})  &  =-\int\frac{d\mathbf{q}%
}{(2\pi)^{2}}\frac{N(\mathbf{q}+\mathbf{k}_{-}/2)-N(\mathbf{q}-\mathbf{k}%
_{-}/2)}{\xi_{\mathbf{q}+\mathbf{k}_{-}/2}-\xi_{\mathbf{q}-\mathbf{k}_{-}/2}%
}\tilde{V}_{2D}(\mathbf{k}_{-})\tilde{V^{\prime}}_{++}(\mathbf{q}%
+\mathbf{k}_{+}/2),\label{dVc}\\
\delta V_{d}(\mathbf{k},\mathbf{k}^{\prime})  &  =-\int\frac{d\mathbf{q}%
}{(2\pi)^{2}}\frac{N(\mathbf{q}+\mathbf{k}_{+}/2)-N(\mathbf{q}-\mathbf{k}%
_{+}/2)}{\xi_{\mathbf{q}+\mathbf{k}_{+}/2}-\xi_{\mathbf{q}-\mathbf{k}_{+}/2}%
}\tilde{V}_{2D}(\mathbf{q}-\mathbf{k}_{-}/2)\tilde{V}_{2D}%
(\mathbf{q}+\mathbf{k}_{-}/2), \label{dVd}%
\end{align}
\end{widetext}
where $\mathbf{k}_{\pm}=\mathbf{k}\pm\mathbf{k}^{\prime}$ and we
keep only momentum dependent part $\tilde{V}_{++}^{\prime}$ of the intralayer
potential because the contributions of the momentum independent part of
$\tilde{V}_{++}$ in $\delta V_{a}$, $\delta V_{b}$, and $\delta V_{c}$ cancel
each other, as it should be. The contribution $\delta V_{a}(\mathbf{k}%
,\mathbf{k}^{\prime})$ can be calculated analytically: for $k=k^{\prime}%
=k_{F}$ we have $k_{-}\leq2k_{F}$ and, hence,
\begin{align}
\delta V_{a}(\mathbf{k},\mathbf{k}^{\prime}) &  =-2\nu_{F}\tilde{V}%
_{2D}(\mathbf{k}_{-})\tilde{V}_{++}^{\prime}(\mathbf{k}_{-})\nonumber\\
&  \approx-\frac{m}{\pi\hbar^{2}}(-\frac{2\pi\hbar^{2}}{m}g\left\vert
\mathbf{k}-\mathbf{k}^{\prime}\right\vert l)^{2}\nonumber\\
&  =-\frac{4\pi\hbar^{2}}{m}(gl)^{2}(\mathbf{k}-\mathbf{k}^{\prime}%
)^{2},\nonumber
\end{align}
while the other three can be computed numerically. The corresponding $s$-wave
contributions are obtained by averaging over the directions of $\mathbf{k}$
and $\mathbf{k}^{\prime}$ (azimuthal angles $\varphi$ and $\varphi^{\prime}$,
respectively):%
\[
\overline{\delta V_{i}}=\left\langle \delta V_{i}(\mathbf{k},\mathbf{k}%
^{\prime})\right\rangle _{\varphi,\varphi^{\prime}}\equiv\int_{0}^{2\pi}%
\frac{d\varphi d\varphi^{\prime}}{(2\pi)^{2}}\delta V_{i}(\mathbf{k}%
,\mathbf{k}^{\prime}),\quad i=a,b,c,d.
\]
The contribution $\overline{\delta V_{i}}$, can be written in the form%
\[
\overline{\delta V_{i}}=\frac{2\pi\hbar^{2}}{m}(gk_{F}l)^{2}\eta_{a},
\]
where $\eta_{a}=-4$ and numerical calculation of the integrals for $\delta
V_{b}=$ $\delta V_{c}$ and $\delta V_{d}$ result in
\[
\eta_{b}=\eta_{c}=1.148,\quad\eta_{d}=0.963.
\]
As a result we obtain%
\begin{equation}
\overline{\delta V}=\frac{2\pi\hbar^{2}}{m}(gk_{F}l)^{2}\sum_{i}\eta
_{i}=-0.741\frac{2\pi\hbar^{2}}{m}(gk_{F}l)^{2}.\label{deltaV_a}%
\end{equation}

In the regime b ($\exp(-1/g^{2})\ll k_{F}l<g<1$) the leading contribution to
the two-particle interlayer scattering is given by the second order Born term,
and the small parameter of the theory characterizing the interlayer scattering
is $\lambda=\nu_{F}\Gamma=-g^{2}/4$. For the intralayer scattering, however,
the leading contribution is still given by the first order Born term $\sim
gk_{F}l$. This is because the intralayer scattering occurs between identical
fermions and, hence, the dominant contribution is the $p$-wave one. The second
order Born contribution in this case is proportional to $(gk_{F}l)^{2}%
\ln(k_{F}l_{0})$ (see Ref. \cite{Sieberer}) or $(gk_{F}l)^{2}\ln(gk_{F}l)$ for
$l_{0}\rightarrow0$ (see Ref. \cite{LevinsenShlyapnikov}) and can be
neglected. As a result, the leading contributions to the effective
interparticle interaction will be given by the same diagram from Fig.
\ref{Fig_deltaV}, in which all interaction $\tilde{V}_{2D}$ lines that connect
fermionic lines belonging to different layer are replaced with the second
order Born scattering amplitude $\Gamma^{(2)}$. It is then easy to see, that
the contribution to $\delta V(\mathbf{k},\mathbf{k}^{\prime})$ comes from the
diagram d and equals ($k=k^{\prime}=k_{F}$)%
\begin{equation}
\delta V(\mathbf{k},\mathbf{k}^{\prime})\approx\nu_{F}\left[  -\frac{2\pi
\hbar^{2}}{m}\frac{g^{2}}{4}\right]  ^{2}=\frac{2\pi\hbar^{2}}{m}\left[
\frac{g^{2}}{4}\right]  ^{2}=\nu_{F}^{-1}\lambda^{2}.\label{deltaV_b}%
\end{equation}

The interlayer scattering amplitude in the regime c ($\exp(-1/g^{2})\lesssim
kl\ll g<1$) is $\Gamma(E,\mathbf{k},\mathbf{k}^{\prime})\approx(4\pi\hbar
^{2}/m)(\ln(E_{b}/E)+i\pi)^{-1}\approx(4\pi\hbar^{2}/m)\ln^{-1}(E_{b}/E)$,
similar to the scattering amplitude for a short-range potential. The
corresponding small parameter is simply $\lambda=2/\ln(E_{b}/\varepsilon_{F}%
)$. (Note the conditions $\operatorname{Re}\lambda<0$ and $\left\vert
\lambda\right\vert <1$ requires $\varepsilon_{F}>E_{b}$.) Arguments, similar
to those given for the regime b, lead us to the conclusion that the leading
many-body contribution to the effective interparticle interaction is given by
the diagram d in Fig. \ref{Fig_deltaV}, in which the interaction lines are
replaced with the scattering amplitude (see analogous considerations in Ref.
\cite{PetrovBaranovShlyapnikov}):%
\begin{equation}
\delta V(\mathbf{k},\mathbf{k}^{\prime})\approx\frac{2\pi\hbar^{2}}{m}\frac
{4}{\ln^{2}(E_{b}/\varepsilon_{F})}=\nu_{F}^{-1}\lambda^{2}.\label{deltaV_c}%
\end{equation}

Note that the leading many-body contribution to the effective interparticle
interaction in the regimes b and c can be written as%
\begin{equation}
\delta V(\mathbf{k},\mathbf{k}^{\prime})=\nu_{F}^{-1}\lambda^{2},
\label{Gamma_bc}%
\end{equation}
which is independent on the directions of $\mathbf{k}$ and $\mathbf{k}%
^{\prime}$ and, hence, coincide with its $s$-wave component, $\overline{\delta
V}=\nu_{F}^{-1}\lambda^{2}$.

\section{Critical temperature in the dilute limit}

We now proceed with the solution of the gap equation (\ref{GapEq}) in the
dilute limit $kl<1$. As we have already pointed out, the order parameter has
the $s$-wave symmetry, $\Delta(\mathbf{k})=\Delta(k)$, and, therefore, it is
convenient to work with the gap equation projected to the $s$-wave channel:%
\begin{align}
\Delta(k) =  &  -\frac{m_{\ast}}{m}\int_{0}^{\infty}\frac{k^{\prime}%
dk^{\prime}}{2\pi}\Gamma_{s}(2\mu,k,k^{\prime})\nonumber\\
&  \quad\left[  \frac{\tanh(\xi_{k^{\prime}}/2T_{c})}{2\xi_{k^{\prime}}}%
+\frac{1}{2\mu-\hbar^{2}k^{\prime2}/m+i0}\right]  \Delta(k^{\prime
})\nonumber\\
&  -\int_{0}^{\Lambda k_{F}}\frac{k^{\prime}dk^{\prime}}{2\pi}\overline{\delta
V}\frac{\tanh(\xi_{k^{\prime}}/2T_{c})}{2\xi_{k^{\prime}}}\Delta(k^{\prime}),
\label{GapEq_s}%
\end{align}
where $\Gamma_{s}(2\mu,k,k^{\prime})$ is the $s$-wave component of the vertex
function for the interlayer scattering, see Eq. (\ref{def_Gamma_s}).

Note, that following our previous discussion, the combination $(m_{\ast
}/m)\Gamma_{s}(2\mu,k,k^{\prime})$ in Eq. (\ref{GapEq_s}) has to be calculated
to second order in the small parameter, and the second order term has to be
taken at the Fermi surface ($k=k^{\prime}=k_{F}$), similar to the
$\overline{\delta V}$ contribution. All these second order terms can be
treated perturbatively.

\subsection{BCS approach}

In the first order in the small parameter Eq. (\ref{GapEq_s}) corresponds to
the BCS gap equation%
\begin{align}
\Delta(k)=  &  -\int_{0}^{\infty}\frac{k^{\prime}dk^{\prime}}{2\pi}\Gamma
_{s}(2\mu,k,k^{\prime})\\
&  \left[  \frac{\tanh(\xi_{k^{\prime}}/2T_{c})}{2\xi_{k^{\prime}}}+\frac
{1}{2\mu-\hbar^{2}k^{\prime2}/m+i0}\right]  \Delta(k^{\prime}).
\nonumber\label{gapBCS_s}%
\end{align}
In order to solve this equation, we rewrite it in the form%
\begin{equation}
\Delta(\xi)=-\int_{-\mu}^{\infty}d\xi^{\prime}R(\xi,\xi^{\prime})\left[
\frac{\tanh(\xi_{k^{\prime}}/2T_{c})}{2\xi_{k^{\prime}}}-\frac{1}%
{2\xi_{k^{\prime}}-i0}\right]  \Delta(\xi^{\prime}), \label{gap_eq1}%
\end{equation}
where $R(\xi,\xi^{\prime})=\nu_{F}\Gamma_{s}(2\mu,k_{\xi},k_{\xi^{\prime}%
}^{\prime})$ with $\xi=\hbar^{2}k^{2}/2m-\mu$, $k_{\xi}=\hbar^{-1}\sqrt
{2m(\xi+\mu)}$, and $k_{\xi^{\prime}}^{\prime}=\hbar^{-1}\sqrt{2m(\xi^{\prime
}+\mu)}$.

We then introduce a characteristic energy $\omega$, which is of the order of
the Fermi energy and, on the other hand, is much larger than the critical
temperature, $\omega\gg T_{c}$, and divide the integral over $\xi^{\prime}$ in
Eq. (\ref{gap_eq1}) into three parts: (a) the integration of $R(\xi
,0)\Delta(0)$ from $-\omega$ to $\omega$, (b) the integration of $R(\xi
,\xi^{\prime})\Delta(\xi^{\prime})-R(\xi,0)\Delta(0)$ from $-\omega$ to
$\omega$, and (c) the integration of $R(\xi,\xi^{\prime})\Delta(\xi^{\prime})$
from $-\mu$ to $-\omega$ and from $\omega$ to $\infty$. In the part (a) we use
the asymptotic formula%
\[
\int_{-\omega}^{\omega}d\xi^{\prime}\frac{\tanh(\xi_{k^{\prime}}/2T_{c})}%
{2\xi_{k^{\prime}}}\approx\ln\frac{2\exp(\gamma)\omega}{\pi T_{c}},
\]
while in parts (b) and (c) we replace $\tanh(\xi_{k^{\prime}}/2T_{c})$ by the
step function (omitting the unimportant contribution from a narrow interval
$\left\vert \xi^{\prime}\right\vert \lesssim T_{c}\ll\omega$) and integrate by
parts. Eq. (\ref{gap_eq1}) then takes the form%
\begin{align}
\Delta(\xi) =  &  -\left[  \ln\frac{2\exp(\gamma)\omega}{\pi T_{c}}-i\frac
{\pi}{2}\right]  R(\xi,0)\Delta(0)\nonumber\\
&  -\ln\frac{\mu}{\omega}R(\xi,-\mu)\Delta(-\mu)\nonumber\\
&  -\int_{-\mu}^{0}d\xi^{\prime}\ln\left\vert \frac{\xi^{\prime}}{\omega
}\right\vert \frac{d}{d\xi^{\prime}}\left[  R(\xi,\xi^{\prime})\Delta
(\xi^{\prime})\right]  , \label{gap_eq2}%
\end{align}
where the first term comes from the part (a).

It is easy to see that the first term is larger than the second and the third
ones by a factor $\ln[2\exp(\gamma)\omega/\pi T_{c}]$, and, therefore, the
last two terms contribute only to the preexponential factor in the expression
for the critical temperature. In order to solve Eq. (\ref{gap_eq2}), we choose
$\omega$ such that%
\begin{equation}
\ln\frac{\mu}{\omega}R(0,-\mu)\Delta(-\mu)+\int_{-\mu}^{0}d\xi^{\prime}%
\ln\left\vert \frac{\xi^{\prime}}{\omega}\right\vert \frac{d}{d\xi^{\prime}%
}\left[  R(0,\xi^{\prime})\Delta(\xi^{\prime})\right]  =0 \label{eq_omega}%
\end{equation}
and, putting $\xi=0$ in (\ref{gap_eq2}), we obtain the following equation to
finding the critical temperature:%
\begin{equation}
\Delta(0)=-\left[  \ln\frac{2\exp(\gamma)\omega}{\pi T_{c}}-i\frac{\pi}%
{2}\right]  R(0,0)\Delta(0). \label{Eq._Tc_BCS}%
\end{equation}
It follows from this equation that
\begin{equation}
\ln\frac{2\exp(\gamma)\omega}{\pi T_{c}}-i\frac{\pi}{2}=-\frac{1}{R(0,0)},
\label{lnTc}%
\end{equation}
and, therefore, after using Eq. (\ref{ImGamma_s_approx}),%
\begin{equation}
T_{c}^{\mathrm{BCS}}=\frac{2\exp(\gamma)}{\pi}\omega\exp\left[  \frac
{1}{R^{\prime}(0,0)}\right]  , \label{Tc_omega}%
\end{equation}
where $R^{\prime}$ is the real part of $R$, $R^{\prime}=\operatorname{Re}R$,
such that $R=R^{\prime}+iR^{\prime\prime}$, and $R^{\prime\prime}\approx
-(\pi/2)(R^{\prime})^{2}$, according to Eq. (\ref{ImGamma_s_approx}).

The value of the energy $\omega$ can be obtained from Eq. (\ref{eq_omega}):%
\begin{align}
\ln\omega=  &  \ln\mu\frac{R(0,-\mu)\Delta(-\mu)}{R(0,0)\Delta(0)}+\frac
{1}{R(0,0)\Delta(0)}\nonumber\label{ln_omega1}\\
&  \qquad\int_{-\mu}^{0}d\xi^{\prime}\ln\left\vert \xi^{\prime}\right\vert
\frac{d}{d\xi^{\prime}}\left[  R(0,\xi^{\prime})\Delta(\xi^{\prime})\right] \\
=  &  \ln\mu+\frac{1}{R(0,0)\Delta(0)}\int_{-\mu}^{0}d\xi^{\prime}\ln
\frac{\left\vert \xi^{\prime}\right\vert }{\mu}\frac{d}{d\xi^{\prime}}\left[
R(0,\xi^{\prime})\Delta(\xi^{\prime})\right]  .\nonumber
\end{align}
Substituting of Eqs. (\ref{lnTc}) and (\ref{ln_omega}) into Eq. (\ref{gap_eq2}%
) results then in the equation for the order parameter%
\begin{align}
\Delta(\xi)=  &  \frac{R(\xi,0)}{R(0,0)}\Delta(0)+\int_{-\mu}^{0}d\xi^{\prime
}\ln\left\vert \frac{\xi^{\prime}}{\mu}\right\vert \frac{d}{d\xi^{\prime}%
}\nonumber\label{eq_order_prameter}\\
&  \quad\left\{  \left[  \frac{R(\xi,0)}{R(0,0)}R(0,\xi^{\prime})-R(\xi
,\xi^{\prime})\right]  \Delta(\xi^{\prime})\right\}  ,
\end{align}
in which the second term is proportional to the small parameter and, hence,
can be considered as a perturbation. Therefore, to leading order in the small
parameter, the solution of Eq. (\ref{eq_order_prameter}) reads%
\begin{equation}
\Delta(\xi)\approx\frac{R(\xi,0)}{R(0,0)}\Delta(0). \label{order_parameter}%
\end{equation}
Substituting this expression into Eq. (\ref{ln_omega1}), we obtain%
\begin{equation}
\ln\omega=\ln\mu+\frac{1}{R(0,0)^{2}}\int_{-\mu}^{0}d\xi^{\prime}\ln
\frac{\left\vert \xi^{\prime}\right\vert }{\mu}\frac{d}{d\xi^{\prime}}\left[
R(0,\xi^{\prime})R(\xi^{\prime},0)\right]  , \label{ln_omega}%
\end{equation}
where we can replace all functions $R$ with their real parts $R^{\prime}$, see
Eq. (\ref{ImGamma_s_approx}). This expression, together with Eq.
(\ref{Tc_omega}), provide the answer for the critical temperature in the BCS approach.

\subsection{Critical temperature in the many-body system}

Following our previous discussion, one should take into account only those
many-body contribution that appear in combination with the large logarithm
$\ln(\varepsilon_{F}/T_{c})$ originating from the momenta close to the Fermi
momenta. Therefore, in view of the many-body contributions, Eq.
(\ref{Eq._Tc_BCS}) can be written as

\begin{align}
\Delta(0)  &  =-\frac{m_{\ast}}{m}\left[  \ln\frac{2e^\gamma\omega}{\pi
T_{c}}-i\frac{\pi}{2}\right]  R(0,0)\Delta(0)\nonumber\\
&\quad-\ln\frac{2e^\gamma\omega
}{\pi T_{c}}\nu_{F}\overline{\delta V}\Delta(0) \nonumber\\
&  \approx-\left[  \frac{m_{\ast}}{m}R(0,0)+\nu_{F}\overline{\delta V}\right]
\left[  \ln\frac{2e^\gamma\omega}{\pi T_{c}}-i\frac{\pi}{2}\right]\nonumber
\Delta(0).
\end{align}
Therefore,
\begin{align}
&\ln\frac{2e^\gamma\omega}{\pi T_{c}}-i\frac{\pi}{2}    =-\frac{1}%
{m_{\ast}R(0,0)/m+\nu_{F}\overline{\delta V}}\nonumber\\
&\approx-\frac{1}{m_{\ast}R^{\prime}(0,0)/m+iR^{\prime\prime}(0,0)+\nu_{F}%
\overline{\delta V}} \nonumber\\
&  \approx-\frac{1}{R^{\prime}(0,0)}+\frac{\left(  \frac{m_{\ast}}{m}-1\right)  R^{\prime}(0,0)+\nu_{F}\overline{\delta
V}+iR^{\prime\prime}(0,0)}{R^{\prime}(0,0)^{2}} \nonumber\\
&  \approx-\frac{1}{R^{\prime}(0,0)}+\frac{1}{R^{\prime}(0,0)}\left(
\frac{m_{\ast}}{m}-1\right)  +\frac{\nu_{F}\overline{\delta V}}{R^{\prime
}(0,0)^{2}}-i\frac{\pi}{2}. \nonumber
\end{align}
As a result, for the critical temperature we obtain %
\begin{align}
T_{c}  &  =\frac{2e^\gamma \omega e^{1/R^{\prime}(0,0)}}{\pi}\exp\left[  -\frac{m_{\ast}/m-1}{R^{\prime
}(0,0)} -\frac{\nu_{F}\overline{\delta V}%
}{R^{\prime}(0,0)^{2}}\right]  
\nonumber\\
&  =T_{c}^{\mathrm{BCS}}\exp\left[  -\frac{m_{\ast}/m-1}{R^{\prime}(0,0)}  -\frac{\nu_{F}\overline{\delta V}}{R^{\prime
}(0,0)^{2}}\right]\label{Tc},
\end{align}%
where $\omega$ and $m_{\ast}$ are given by Eqs. (\ref{ln_omega}) and
(\ref{effective_mass}), respectively. The specific expression for the
many-body contribution to the effective interparticle interaction
$\overline{\delta V}$ depends on the regime of scattering, see Eqs.
(\ref{deltaV_a}), (\ref{deltaV_b}) and (\ref{deltaV_c}).

We now analyze the expression (\ref{Tc}) for the critical temperature for
different regimes of scattering:

\emph{Regime a: }For $g<k_{F}l<1$, we have%
\begin{align}
R^{\prime}(0,0)  &  \approx\nu_{F}\Gamma_{s}^{(1)}(k_{F})-\frac{g^{2}}%
{4}\left\{  1-2(k_{F}l)^{2}\left[  5.4+3\ln(k_{F}l)\right]  \right\}\nonumber\\
&\quad
-\frac{4g^{3}}{15}\label{a1}\\
&  \approx-gk_{F}l\frac{4}{\pi}(1-\frac{\pi}{2}k_{F}l)\nonumber\\
&\quad-\frac{g^{2}}{4}\left\{
1-2(k_{F}l)^{2}\left[  5.4+3\ln(k_{F}l)\right]  \right\}  -\frac{4g^{3}}%
{15},\nonumber\\
\frac{m_{\ast}}{m}-1  &  \approx-\frac{4}{3\pi}gk_{F}l,\label{a2}\\
\nu_{F}\overline{\delta V}  &  \approx-0.741(gk_{F}l)^{2}, \label{a3}%
\end{align}%
where we expand $\nu_{F}\Gamma_{s}^{(1)}(k_{F})$ in the expression for
$R^{\prime}(0,0)$ up to the second order in powers of $k_{F}l$, and keep only
those terms that give contributions up to order unity in the expression for
the critical temperature. For the calculation of $\omega$, see Eq.
(\ref{ln_omega}), it is sufficient to take $R(0,0)$ in the form $R(0,0)=\nu
_{F}\left\langle \tilde{V}_{2D}(\mathbf{k}-\mathbf{k}^{\prime})\right\rangle
_{\varphi,\varphi^{\prime}}$. The resulting integration can be performed in
the same way as for the integral $I_{2}$ from Appendix D, and we obtain%
\[
\omega=\mu\exp\left[  -0.697\left(  \frac{\pi}{4}\right)  ^{2}\right]
=0.651\mu.
\]
With the help of Eqs. (\ref{a1})-(\ref{a3}) we can write (within the accepted
accuracy)
\begin{align}
\frac{1}{R^{\prime}(0,0)}  &  \approx-\left[  \nu_{F}\left\vert \Gamma
_{s}^{(1)}(k_{F})\right\vert +\frac{g^{2}}{4}+\frac{4g^{3}}{15}\right]
^{-1}\nonumber\\
&\quad-\frac{1}{2}\left(  \frac{\pi}{4}\right)  ^{2}\left[  5.4+3\ln
(k_{F}l)\right] \\
&  \approx-\left[  gk_{F}l\frac{4}{\pi}(1-\frac{\pi}{2}k_{F}l)+\frac{g^{2}}%
{4}+\frac{4g^{3}}{15}\right]  ^{-1}\nonumber\\
&\quad-1.17-0.925\ln(k_{F}l),
\\
\frac{ m_{\ast}/m-1}{R^{\prime}(0,0)}  &\approx\frac
{1}{3},\nonumber\\
\frac{\nu_{F}\overline{\delta V}}{[R^{\prime}(0,0)]^{2}}&\approx-0.741\left(
\frac{\pi}{4}\right)  ^{2}=-0.457.\nonumber
\end{align}

From Eq. (\ref{Tc}) we now obtain the final expression for the critical
temperature for the BCS pairing in the regime a of interparticle
scattering
\begin{align}\label{Tca}
T_{c,a} &  =\frac{2e^\gamma}{\pi}0.651\mu\exp\left\{  -\left[  \nu
_{F}\left\vert \Gamma_{s}^{(1)}(k_{F})\right\vert +\frac{g^{2}}{4}%
+\frac{4g^{3}}{15}\right]  ^{-1}\right.\nonumber\\
& \qquad \qquad \qquad\left.-1.17-0.925\ln(k_{F}l)+\frac{-1}{3}+0.457\right\}  \nonumber\\
&  \approx0.259\mu(k_{F}l)^{-0.925}\\
&\quad\exp\left\{  -\left[  gk_{F}l\left(\frac{4}{\pi
}-\frac{k_{F}l}{2}\right)+\frac{g^{2}}{4}+\frac{4g^{3}}{15}\right]
^{-1}\right\} \nonumber,
\end{align}
where for $k_{F}l\lesssim0.2$ one has $\nu_{F}\Gamma_{s}%
^{(1)}(k_{F})\approx gk_{F}l\frac{4}{\pi}(1-\frac{\pi}{2}k_{F}l)$. The
comparison of this results with the one obtained in Ref. \cite{GoraSantos} in
the BCS approach shows that the many-body effects result in a larger (by a
factor of two) numerical prefactor. This is the effect of the competition of
decreasing of the critical temperature because of the smaller effective mass
and increasing $T_{c}$ because of the attractive many-body contribution to the
interparticle interaction. In addition, Eq. (\ref{Tca}) contains an extra term
$4g^{3}/15$ in the denominator in the exponent, originating from the third
order contribution in the particle scattering amplitude. This term is smaller
than the other two. However, one needs a stronger condition, namely
$g<(k_{F}l)^{3/2}$, to neglect this term. This is because being expanded, this
it results in the contribution $\sim g^{3}/(k_{F}l)^{3}=$ $(g/k_{F}%
l)^{2}(k_{F}l)^{-1}$, which is small only under the stronger condition. Note
that this term also leads to the higher critical temperature the in the BCS approach.

The order parameter, Eq. (\ref{order_parameter}), in this regime has the form
($k^{\prime}=k_{F}$)%
\begin{align*}
\Delta(k)  &  \sim g\left\langle \left\vert \mathbf{k}-\mathbf{k}^{\prime
}\right\vert l\exp(-\left\vert \mathbf{k}-\mathbf{k}^{\prime}\right\vert
l)\right\rangle _{\varphi,\varphi^{\prime}}+\frac{g^{2}}{4}\\
&  \approx g\left\langle \left\vert \mathbf{k}-\mathbf{k}^{\prime}\right\vert
l(1-\left\vert \mathbf{k}-\mathbf{k}^{\prime}\right\vert l)\right\rangle
_{\varphi,\varphi^{\prime}}+\frac{g^{2}}{4}\\
&  =g\left\{  \frac{2}{\pi}(k+k_{F})l\ \mathrm{E}[\frac{4kk_{F}}{(k+k_{F}%
)^{2}}]-(k^{2}+k_{F}^{2})l^{2}\right\}  +\frac{g^{2}}{4},
\end{align*}
where $\mathrm{E}(z)$ is the complete elliptic integral and we assume
$kl,\ k_{F}l\lesssim0.2$ to ensure the reasonable accuracy of the truncated expansion.

\emph{Regime b: }In the regime b, $k_{F}l<g<1$, we have%
\begin{align}
R^{\prime}(0,0)\approx &  -\frac{g^{2}}{4}-gk_{F}l\frac{4}{\pi}(1-\frac{\pi
}{2}k_{F}l)-\frac{4g^{3}}{15}\nonumber\\
&  -\frac{g^{4}}{32}\left[  \ln(4\hbar^{2}/m\mu l^{2})+\frac{7}{2}%
-2\gamma\right]  ,\label{b1}\\
\approx &  \frac{2}{\ln(E_{b}/\mu)}-gk_{F}l\frac{4}{\pi}(1-\frac{\pi}{2}%
k_{F}l),\nonumber\\
\frac{m_{\ast}}{m}-1\approx &  -\frac{4}{3\pi}gk_{F}l,\label{b2}\\
\nu_{F}\overline{\delta V}\approx &  \left[  \frac{g^{2}}{4}\right]  ^{2}.
\label{b3}%
\end{align}
The leading order contribution in $R^{\prime}(0,0)$ is momentum and energy
independent and, therefore, we have $\omega=\mu$. From Eqs. (\ref{b1}%
)-(\ref{b3}) we obtain%
\begin{align}
\frac{1}{R^{\prime}(0,0)}\approx &  -\left[  \frac{g^{2}}{4}+gk_{F}l\frac
{4}{\pi}(1-\frac{\pi}{2}k_{F}l)+\frac{4g^{3}}{15}\right]  ^{-1}\nonumber\\
&  +\frac{1}{2}\left[  \ln\left(\frac{4\hbar^{2}}{m\mu l^{2}}\right)+\frac{7}{2}-2\gamma\right]
,\nonumber\\
\frac{m_\star/m-1}{R^{\prime}(0,0)}  \approx &
~\frac{14}{3\pi}\frac{k_{F}l}{g}\ll1,\nonumber\\
\frac{\nu_{F}\overline{\delta V}}{[R^{\prime}(0,0)]^{2}}\approx &  ~1.
\end{align}
Note that the many-body contribution to the effective mass is negligible
because the leading term in the scattering amplitude is momentum and energy
independent. From Eq. (\ref{Tc}) we then obtain%
\begin{align}\label{Tcb}
&T_{c,b}    =\frac{2e^\gamma}{\pi}\mu\exp\left\{  -\left[  \frac{g^{2}}%
{4}+gk_{F}l\frac{4}{\pi}(1-\frac{\pi}{2}k_{F}l)+\frac{4g^{3}}{15}\right]
^{-1}\right. \nonumber\\
&  \qquad \qquad \quad\left.  +\frac{1}{2}\left[  \ln\left(\frac{4\hbar^{2}}{m\mu l^{2}}\right)+\frac{7}{2}-2\gamma\right]  -1\right\}\\
&  =\frac{4e^{3/4}}{\pi}\sqrt{\frac{\mu\hbar^{2}}{ml^{2}}}\exp\left\{
-\left[  \frac{g^{2}}{4}+gk_{F}l\left(\frac{4}{\pi}-2k_{F}l\right)+\frac{4g^{3}}{15}\right]  ^{-1}\right\}\nonumber\\
&  =2.7\sqrt{\frac{\mu\hbar^{2}}{ml^{2}}}\exp\left[  -\frac{1}{g^{2}/4+4gk_{F}l/\pi-2gk_F^2l^2+4g^{3}/15}\right].\nonumber%
\end{align}
Furthermore, we note that the exponent in this expression coincides with that
in Eq. (\ref{Tca}) and, following the same arguments as in Eq. (\ref{Tca}), we
keep some higher terms in the denominator in the exponent.

\emph{Regime c: }Finally, for $\exp(-1/g^{2})\lesssim k_{F}l\ll g<1$,%
\begin{align*}
R^{\prime}(0,0)  &  \approx\frac{2}{\ln(E_{b}/\mu)},\\
\frac{m_{\ast}}{m}-1  &  \approx\frac{4}{3\pi}gk_{F}l\ll1,\\
\nu_{F}\overline{\delta V}  &  \approx\left[  \frac{2}{\ln(E_{b}/\mu)}\right]
^{2},
\end{align*}
where $E_{b}$ is given by Eq. (\ref{Eb}). Therefore, taking into account that,
similar to the regime b, $\omega=\mu$, we obtain%
\begin{align}
T_{c,c}  &  =\frac{2e^\gamma}{\pi}\mu\exp\left\{  \frac{1}{2}\ln(E_{b}%
/\mu)-1\right\} \nonumber\\
&  =\frac{2\exp(\gamma-1)}{\pi}\sqrt{\mu E_{b}}=0.42\sqrt{\mu E_{b}}.
\label{Tcc}%
\end{align}
This expression is completely analogous to the critical temperature in a
two-component 2D Fermi gas with a short-range interparticle interaction (see
Ref. \cite{PetrovBaranovShlyapnikov}), as it should be in this regime.

Note that within the accepted accuracy, both expressions (\ref{Tcb}) and
(\ref{Tcc}) for the critical temperature in the regimes b and c can be written
in the form%
\begin{align}
T_{c,bc}  &  =\frac{2 e^{\gamma}}{\pi e} \mu\exp\left\{  \left[  \frac{2}%
{\ln(E_{b}/\mu)}-\frac{gk_{F}l}{\pi/4}\left(1-\frac{k_{F}l}{2/\pi}\right)\right]
^{-1}\right\} \label{Tcbc}\\
&  =0.42\mu\exp\left\{  \left[  \frac{2}{\ln(E_{b}/\mu)}-gk_{F}l\frac{4}{\pi
}(1-\frac{\pi}{2}k_{F}l)\right]  ^{-1}\right\}  .\nonumber
\end{align}
In both these regimes, the order parameter is to the leading order momentum
independent, $\Delta(k)\sim\mathrm{const}$.

Eqs. (\ref{Tca}) and (\ref{Tcbc}) provides the answer for the critical
temperature of the BCS transition in a dilute bilayer dipolar gas. The
corresponding values of $T_{c}$ calculated according to these formulae are
small, $T_{c}\lesssim10^{-3}\div10^{-2}\mu$, because the exponent in Eqs.
(\ref{Tca}) and (\ref{Tcbc}) contains the inverse of the product (or square)
of the small parameters of the problem. For example, even for $k_{F}l=0.5$ and
$g=0.45<k_{F}l$ one has $T_{c}\approx10^{-2}\mu$. This makes an experimental
realization of the superfluid state very challenging. The situation is more
promising in the dense case.

\section{Critical temperature in the dense limit}

We assume now that $k_{F}l\gtrsim1$ and $g\ll1$ such that $gk_{F}l=k_{F}%
a_{d}<1$(and $k_{F}l_{0}\ll1$). The condition $gk_{F}l<1$ ensures the validity
of the perturbative expansion in powers of the interlayer interaction,
although the mean interparticle interaction is comparable or larger that the
range of the potential $l$. For this reason there is no need to renormalize
the gap equation: two colliding particles are no more well-separated from the
rest of the system, but many-particles collisions are still well-controlled by
the small parameter $a_{d}k_{F}=gk_{F}l<1$. With the account of the many-body
effects, the gap equation for the pairing in the $s$-wave channel reads%
\begin{equation}
\Delta(k)=-\frac{m_{\ast}}{m}\int_{0}^{\infty}\frac{k^{\prime}dk^{\prime}%
}{2\pi}V_{\mathrm{eff},s}(k,k^{\prime})\frac{\tanh(\xi_{k^{\prime}}/2T_{c}%
)}{2\xi_{k^{\prime}}}\Delta(k^{\prime}), \label{gap_eq_dense}%
\end{equation}
where $V_{\mathrm{eff},s}(k,k^{\prime})$ is the effective interparticle
interaction in the $s$-wave channel,
\begin{align*}
V_{\mathrm{eff},s}(k,k^{\prime})  &  =\left\langle \tilde{V}_{2D}%
(\mathbf{k}-\mathbf{k}^{\prime})+\delta V(\mathbf{k},\mathbf{k}^{\prime
})\right\rangle _{\varphi,\varphi^{\prime}}\\
&  =\Gamma_{s}^{(1)}(k,k^{\prime})+\left\langle \delta V(\mathbf{k}%
,\mathbf{k}^{\prime})\right\rangle _{\varphi,\varphi^{\prime}}.
\end{align*}
Here%
\[
\Gamma_{s}^{(1)}(k,k^{\prime})=\left\langle \tilde{V}_{2D}(\mathbf{k}%
-\mathbf{k}^{\prime})\right\rangle _{\varphi,\varphi^{\prime}}%
\]
and, as before, all many-body corrections have to be taken at the Fermi
surface (assuming $T_{c}\ll\mu$). Using the same arguments as in the previous
Section, we can argue that the critical temperature $T_{c}$ is related to the
critical temperature in the BCS-approach $T_{c}^{\mathrm{BCS}}$ (with no
many-body contributions) as%

\begin{equation}
T_{c}=T_{c}^{\mathrm{BCS}}\exp\left[  -\frac{m_{\ast}/m-1}{R^{\prime}(0,0)}  -\frac{\nu_{F}\overline{\delta V}}{[R^{\prime
}(0,0)]^{2}}\right]  , \label{Tc_vs_TcBCS}%
\end{equation}
where $R^{\prime}(0,0)=\nu_{F}\Gamma_{s}^{(1)}(k_{F},k_{F})$ and
$T_{c}^{\mathrm{BCS}}$ is determined by the BCS gap equation%
\begin{align}
\Delta(k)  &  =-\int_{0}^{\infty}\frac{k^{\prime}dk^{\prime}}{2\pi
}\left\langle \tilde{V}_{2D}(\mathbf{k}-\mathbf{k}^{\prime})\right\rangle
_{\varphi,\varphi^{\prime}}\nonumber\\
&\qquad \qquad \qquad \times\frac{\tanh(\xi_{k^{\prime}}/2T_{c}^{\mathrm{BCS}%
})}{2\xi_{k^{\prime}}}\Delta(k^{\prime})\nonumber\\
&  =-\int_{0}^{\infty}\frac{k^{\prime}dk^{\prime}}{2\pi}\Gamma_{s}%
^{(1)}(k,k^{\prime})\frac{\tanh(\xi_{k^{\prime}}/2T_{c}^{\mathrm{BCS}})}%
{2\xi_{k^{\prime}}}\Delta(k^{\prime}). \label{gap_eq_dense_BCS}%
\end{align}
The solution of this equation is actually given by Eqs. (\ref{Tc_omega}) and
(\ref{ln_omega}): although we are dealing with the non-renormalized gap
equation, the renormalization can still be performed as a formal trick. [Note,
that within the accepted accuracy, $R^{\prime}(0,0)$ in Eq. (\ref{Tc_omega})
has to be calculated up to the second Born approximation, while only up to the
first order in Eq. (\ref{ln_omega}).] Nevertheless, we give here an
alternative solution of the gap equation that follows the lines of Ref.
\cite{KhodelKhodelClark} and avoid the renormalization.

\subsection{BCS approach}

We rewrite Eq. (\ref{gap_eq_dense_BCS}) in the form%
\begin{equation}
\Delta(\xi)=-\int_{-\mu}^{\infty}d\xi^{\prime}\frac{\tanh(\xi^\prime
/2T_{c}^{\mathrm{BCS}})}{2\xi^{\prime}}R(\xi,\xi^{\prime})\Delta(\xi^{\prime
}), \label{gap_eq_dense_BCS_ksi}%
\end{equation}
where $\xi=\hbar^{2}(k^{2}-k_{F}^{2})/2m$,
\begin{align}
R(\xi,\xi^{\prime})   &=\nu_{F}\Gamma_{s}^{(1)}(k,k^{\prime})=gl\int_{0}^{\pi}\frac{d\varphi}{\pi}e^{-l\sqrt{k^{2}+k^{\prime2}-2kk^{\prime}\cos\varphi}}\nonumber\\
&\qquad\qquad\times\sqrt{k^{2}+k^{\prime2}-2kk^{\prime
}\cos\varphi},
\end{align}
and, following the method of Ref. \cite{KhodelKhodelClark},
decompose the interaction function $R(\xi,\xi^{\prime})$ into a separable part
and a remainder $r(\xi,\xi^{\prime})$ that vanishes when either argument is on
the Fermi surface:
\begin{equation}
R(\xi,\xi^{\prime})=R(0,0)v(\xi)v(\xi^{\prime})+r(\xi,\xi^{\prime})
\label{R_decomposition}%
\end{equation}
with $v(\xi)=R(\xi,0)/R(0,0)=R(0,\xi)/R(0,0)$ and $r(\xi,0)=r(0,\xi^{\prime
})=0$. Note that $v(0)=1$ and $v(\xi)$ decays exponentially at large momenta
$kl\gg1$, i.e. $ml^{2}\xi/\hbar^{2}\gg1$. Eq. (\ref{gap_eq_dense_BCS_ksi})
then takes the form:
\begin{align}
\Delta(\xi)=  &  -R(0,0)v(\xi)\int_{-\mu}^{\infty}d\xi^{\prime}\frac{\tanh
(\xi^\prime/2T_{c}^{\mathrm{BCS}})}{2\xi^{\prime}}v(\xi^{\prime})\Delta
(\xi^{\prime})\nonumber\\
&  -\int_{-\mu}^{\infty}d\xi^{\prime}\frac{\tanh(\xi^\prime/2T_{c}%
^{\mathrm{BCS}})}{2\xi^{\prime}}r(\xi,\xi^{\prime})\Delta(\xi^{\prime}).
\label{gap_eq_dense_BCS1}%
\end{align}
For $\xi=0$ this equation reduces to%
\begin{equation}
\Delta(0)=-R(0,0)\int_{-\mu}^{\infty}d\xi^{\prime}\frac{\tanh(\xi^\prime
/2T_{c}^{\mathrm{BCS}})}{2\xi^{\prime}}v(\xi^{\prime})\Delta(\xi^{\prime})
\label{delta0_dense}%
\end{equation}
and, therefore, we can rewrite Eq. (\ref{gap_eq_dense_BCS1}) as follows%
\begin{align}
\Delta(\xi)  &  =v(\xi)\Delta(0)-\int_{-\mu}^{\infty}d\xi^{\prime}\frac
{\tanh(\xi^\prime/2T_{c}^{\mathrm{BCS}})}{2\xi^{\prime}}r(\xi,\xi^{\prime
})\Delta(\xi^{\prime})\nonumber\\
&  \approx v(\xi)\Delta(0)-\int_{-\mu}^{\infty}\frac{d\xi^{\prime}%
}{2\left\vert \xi^{\prime}\right\vert }r(\xi,\xi^{\prime})\Delta(\xi^{\prime
}), \label{delta_xsi_dense1}%
\end{align}
where we replace $\tanh(\xi^\prime/2T_{c}^{\mathrm{BCS}})$ with $\mathrm{sign}%
(\xi^{\prime})$ assuming that $T_{c}\ll\mu$ and neglecting exponentially small
contributions [the integral is converging because $r(\xi,0)=0$]. In Eq.
(\ref{delta0_dense}) we can now single out the large logarithmic contribution
that comes from momenta near the Fermi surface. This can be achieved by
writing $d\xi^{\prime}/\xi^{\prime}=d(\ln\xi^{\prime})$ and integrating by
part with the following result%
\begin{align}
\Delta(0)  &  =-R(0,0)\left\{  \frac{1}{2}v(-\mu)\Delta(-\mu)\ln\mu\right.
\nonumber\\
&  \left.  -\frac{1}{2}\int_{-\mu}^{\infty}d\xi^{\prime}\ln\left\vert
\xi^{\prime}\right\vert \frac{d}{d\xi^{\prime}}\left[  \tanh(\xi^\prime
/2T_{c}^{\mathrm{BCS}})v(\xi^{\prime})\Delta(\xi^{\prime})\right]  \right\}
.\nonumber
\end{align}
After performing the derivative,%
\begin{align}
&  \frac{d}{d\xi^{\prime}}\left[  \tanh(\frac{\xi^\prime}{2T_{c}^{\mathrm{BCS}%
}})v(\xi^{\prime})\Delta(\xi^{\prime})\right]  =\frac{1}{2T_{c}^{\mathrm{BCS}%
}}\frac{1}{\cosh^{2}(\xi^\prime/2T_{c}^{\mathrm{BCS}})}\nonumber\\
&  \qquad\qquad v(\xi^{\prime})\Delta(\xi^{\prime})+\tanh(\frac{\xi^\prime
}{2T_{c}^{\mathrm{BCS}}})\frac{d}{d\xi^{\prime}}\left[  v(\xi^{\prime}%
)\Delta(\xi^{\prime})\right]  ,\nonumber
\end{align}
and using the fact that $1/2T_{c}^{\mathrm{BCS}}\cosh^{2}(\xi^\prime
/2T_{c}^{\mathrm{BCS}})$ is sharply peaked at $\xi^{\prime}=0$, we obtain%
\begin{align}
\Delta(0)  &  =-R(0,0)\left\{  \frac{1}{2}v(-\mu)\Delta(-\mu)\ln\mu+\ln
\frac{2e^\gamma}{\pi T_{c}^{\mathrm{BCS}}}\Delta(0)\right. \nonumber\\
&  \qquad\left.  -\frac{1}{2}\int_{-\mu}^{\infty}d\xi^{\prime}\ln\left\vert
\xi^{\prime}\right\vert \mathrm{sign}(\xi^\prime)\frac{d}{d\xi^{\prime}}\left[
v(\xi^{\prime})\Delta(\xi^{\prime})\right]  \right\} \nonumber\\
&  =-R(0,0)\left\{  \ln\frac{2\mu e^\gamma}{\pi T_{c}^{\mathrm{BCS}}}%
\Delta(0)\right. \nonumber\\
&  \qquad\left.  -\frac{1}{2}\int_{-\mu}^{\infty}d\xi^{\prime}\ln
\frac{\left\vert \xi^{\prime}\right\vert }{\mu}\frac{d}{d\left\vert
\xi^{\prime}\right\vert }\left[  v(\xi^{\prime})\Delta(\xi^{\prime})\right]
\right\}  , \label{delta0_dense1}%
\end{align}
where we replace again $\tanh(\xi^\prime/2T_{c}^{\mathrm{BCS}})$ with
$\mathrm{sign}(\xi^{\prime})$.

The pair of equations (\ref{delta0_dense1}) and (\ref{delta_xsi_dense1}) can
now be solved iteratively because the integrals in both equations provides
small corrections when $k_{F}l\gtrsim1$ (see below). In the leading order, we
have for the order parameter%
$<$%
$\xi^{\prime}$%
\[
\Delta(\xi)\approx v(\xi)\Delta(0).
\]
[Note that the second iteration of Eq. (\ref{delta_xsi_dense1}) with the
explicit expression $r(\xi,\xi^{\prime})=R(\xi,\xi^{\prime})-R(\xi
,0)R(0,\xi^{\prime})/R(0,0)$ can be rewritten in the form of Eq.
(\ref{eq_order_prameter}).] From Eq. (\ref{delta0_dense1}) we then obtain%
\begin{align}
T_{c}^{\mathrm{BCS}}=  &  \frac{2e^\gamma}{\pi}\mu\exp\left\{  -\frac
{1}{2}\int_{-\mu}^{\infty}d\xi^{\prime}\ln\frac{\left\vert \xi^{\prime
}\right\vert }{\mu}\frac{d}{d\left\vert \xi^{\prime}\right\vert }\left[
v(\xi^{\prime})^{2}\right]  \right\} \nonumber\\
&  \exp\left[  \frac{1}{R(0,0)}\right]  \label{Tc_BCS_dense1}%
\end{align}
for the critical temperature in the BCS approach.

To establish the connection with the approach from the previous Section, we
notice that%
\begin{align}
&  -\frac{1}{2}\int_{-\mu}^{\infty}d\xi^{\prime}\ln\frac{\left\vert
\xi^{\prime}\right\vert }{\mu}\frac{d}{d\left\vert \xi^{\prime}\right\vert
}\left[  v(\xi^{\prime})^{2}\right]  =\nonumber\\
&  -\frac{1}{2}\int_{-\mu}^{\infty}d\xi^{\prime}\ln\frac{\left\vert
\xi^{\prime}\right\vert }{\mu}\frac{d}{d\xi^{\prime}}\left[  v(\xi^{\prime
})^{2}\right]  +\int_{-\mu}^{0}d\xi^{\prime}\ln\frac{\left\vert \xi^{\prime
}\right\vert }{\mu}\frac{d}{d\xi^{\prime}}\left[  v(\xi^{\prime})^{2}\right]
.\nonumber
\end{align}
The last term in the right-hand side can now be identified with the
contribution to $\ln\omega$, see Eq. (\ref{ln_omega}), while the first one
with the second order Born contribution to $R^{\prime}(0,0)$.

Before going further, let us discuss the conditions for the iterative approach
to the system of Eqs. (\ref{delta0_dense1}) and (\ref{delta_xsi_dense1}) to be
legitimate. In Eq. (\ref{delta_xsi_dense1}), this requires that the second
term is small and, hence, $\Delta(\xi)/\Delta(0)\approx v(\xi)$. For
$k_{F}l\sim1$, one can see that the relative contribution of the second term
is of the order of $gk_{F}l$. Therefore, the iterative scheme with the result
$\Delta(\xi)\approx v(\xi)\Delta(0)$ is legitimate for $gk_{F}l=a_{d}l<1$. For
a dilute system with $k_{F}l\ll1$, the situation is more subtle. In this case,
the contribution from $\xi^{\prime}\lesssim\mu$ in the second term gives the
relative contribution of the order of $gk_{F}l$, while the regime $\mu
<\xi^{\prime}\lesssim\hbar^{2}/ml^{2}$ results in the relative contribution
$\sim g/k_{F}l$. Therefore, for $g<k_{F}l$ both contributions are small and
the iterative procedure is legitimate. However, for the opposite case
$g>k_{F}l$, the contribution from the second region is large and the iterative
scheme breaks down. The reason for this is the large value of $r(\xi
,\xi^{\prime})$ for $\xi^{\prime}\gg\mu$. The proper iterative procedure in
this case can be developed on the basis of the renormalized gap equation.

\subsection{Critical temperature in the many-body system}

\begin{figure}[ptb]
\begin{center}
\includegraphics[width=.9\columnwidth]{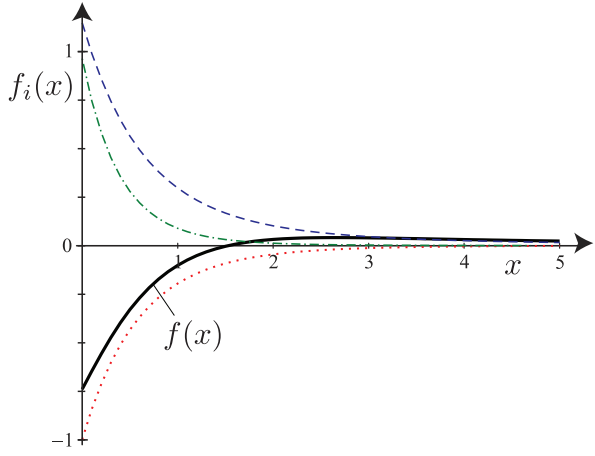}
\end{center}
\caption{The functions $f_a(x)/4$, $f_b(x)=f_c(x)$, $f_d(x)$, and $f(x)$ (dotted, dash-dotted, dashed, and solid lines, respectively).}%
\label{Fig_fabcd}%
\end{figure}

The calculation of the many-body contributions to the critical temperature can
be performed in the same way as in the dilute regime because, although
$k_{F}l\gtrsim1$, we assume that $gk_{F}l=a_{d}k_{F}<1$, and, hence, the
expansion in powers of interparticle interaction is still valid. The
corresponding contributions to the self-energy (effective mass) and
interparticle interaction are shown in Figs. \ref{Fig_Sigma} and
\ref{Fig_deltaV}, and the analytic expressions are given by Eqs.
(\ref{effective_mass}) and (\ref{dVa})-(\ref{dVd}), respectively. Note that
$\delta V_{a}(\mathbf{k},\mathbf{k}^{\prime})$ can be calculated
analytically,
\begin{align*}
\delta V_{a}(\mathbf{k},\mathbf{k}^{\prime}) &  =-2\nu_{F}\tilde{V}%
_{2D}(\mathbf{k}-\mathbf{k}^{\prime})\tilde{V}_{++}^{\prime}(\mathbf{k}%
-\mathbf{k}^{\prime})\\
&  \approx-\frac{4\pi\hbar^{2}}{m}(g\left\vert \mathbf{k}-\mathbf{k}^{\prime
}\right\vert l)^{2}\exp(-\left\vert \mathbf{k}-\mathbf{k}^{\prime}\right\vert
l),
\end{align*}
[as before, we keep only the momentum dependent part of $\tilde{V}%
_{++}(\mathbf{k}-\mathbf{k}^{\prime})$] together with its angular average for
$k=k^{\prime}=k_{F}$,%
\begin{align}
\nu_{F}\delta\overline{V}_{a}= &  -8(gk_{F}l)^{2}\left\{  \mathrm{I}%
_{0}(2k_{F}l)-\frac{1}{2}{}_{0}F_{1}(2;k_{F}^{2}l^{2})\right.  \nonumber\\
&  \qquad\qquad\left.  -\frac{8}{3\pi}(k_{F}l)_{1}F_{2}(2;\frac{3}{2},\frac
{5}{2};k_{F}^{2}l^{2})\right\}  ,\nonumber
\end{align}
where $\mathrm{I}_{0}(z)$ is the modified Bessel function and $_{n}F_{m}%
(a_{1},\ldots,a_{n};b_{1},\ldots,b_{m};z)$ is the hypergeometric function,
while the other three contributions require numerical integrations. We write
these contributions as%
\[
\nu_{F}\delta\overline{V}_{i}=(gk_{F}l)^{2}f_{i}(k_{F}l),\;i=a,b,c,d,
\]
where the functions $f_{i}(x)$ are shown in Fig. \ref{Fig_fabcd}. 
The overall angular averaged contribution to the effective
interaction then reads%
\[
\nu_{F}\delta\overline{V}=(gk_{F}l)^{2}f(k_{F}l),
\]
where the function $f(x)=f_{a}(x)+2f_{b}(x)+f_{c}(x)$ is also shown in Fig.
\ref{Fig_fabcd} (solid line).

\begin{figure}[ptb]
\begin{center}
\includegraphics[width=.8\columnwidth]{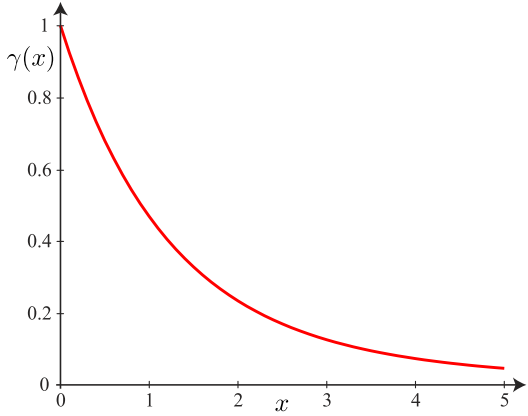}
\end{center}
\caption{The function $\gamma(x)$.}%
\label{Fig_gamma}%
\end{figure}

After writing $R(0,0)$ in the form%
\[
R(0,0)=-\frac{4}{\pi}gk_{F}l\,\gamma(k_{F}l\,),
\]
where%
\[
\gamma(x)=\frac{1}{2}\int_{0}^{\pi}d\varphi\sin(\varphi)e^{-x\sin
(\varphi)}=\frac{\pi}{4}\left[  \mathbf{L}_{-1}(2x)-\mathrm{I}_{1}(2x)\right]
,
\]
see Eq.~\eqref{Gamma1s}, is shown in Fig. \ref{Fig_gamma},  we obtain%
\begin{align}
-\frac{1}{R(0,0)}\left(  \frac{m_{\ast}}{m}-1\right)  -\frac{\nu_{F}%
\overline{\delta V}}{[R(0,0)]^{2}}   =\nonumber\\
 -\frac{1}{3\gamma(k_{F}l\,)}-\left(  \frac{\pi}{4}\right)  ^{2}%
\frac{f(k_{F}l\,)}{\gamma(k_{F}l\,)^{2}}.\label{many_body}%
\end{align}
Finally, after combining together Eqs. (\ref{Tc_BCS_dense1}),
(\ref{Tc_vs_TcBCS}) and (\ref{many_body}), we find%
\begin{align}
T_{c}= &  \frac{2e^{\gamma}}{\pi}\mu\exp\left[  -\frac{1}{16}\frac
{\Omega(k_{F}l)}{\gamma(k_{F}l\,)^{2}}-\frac{1}{3\gamma(k_{F}l\,)}-\left(
\frac{\pi}{4}\right)  ^{2}\frac{f(k_{F}l\,)}{\gamma(k_{F}l\,)^{2}}\right]
\nonumber\label{Tc_dense}\\
&  \exp\left[  -\frac{\pi}{4gk_{F}l\,}\frac{1}{\gamma(k_{F}l\,)}\right]  ,
\end{align}
where%
\[
\Omega(x)=\frac{1}{2}\int_{0}^{\infty}ds\ln\left\vert 1-s^{2}\right\vert
\mathrm{sign}(s-1)\frac{d}{dx}\left[V(s,x)^{2}\right]
\]
and%
\[
V(s,x)=\int_{0}^{\pi}d\varphi\sqrt{1+s^{2}-2s\cos\varphi}e^{-x\sqrt
{1+s^{2}-2s\cos\varphi}}.
\]
The expression (\ref{Tc_dense}) is appropriate for $k_{F}l\gtrsim1$. Based on
the discussion after Eq. (\ref{Tc_BCS_dense1}), we can write the expression
for the critical temperature that will interpolate the behavior for
$k_{F}l\gtrsim1$ and $k_{F}l\lesssim1$:%
\begin{align}
T_{c}= &  \frac{2e^{\gamma}}{\pi}\mu\exp\left[  -\frac{1}{3\gamma(k_{F}%
l\,)}-\left(  \frac{\pi}{4}\right)  ^{2}\frac{f(k_{F}l\,)}{\gamma
(k_{F}l\,)^{2}}\right]  \nonumber\\
&  \exp\left[  -\frac{\pi}{4gk_{F}l\,\gamma(k_{F}l\,)}\frac{1}{1-(4/\pi
)gk_{F}l\,\gamma(k_{F}l\,)\Omega(k_{F}l)}\right]  \nonumber\\
\equiv &  \frac{2e^{\gamma}\mu}{\pi}\tau(g,k_{F}l).\label{Tcinterp}%
\end{align}

\begin{figure}[ptb]
\begin{center}
\includegraphics[width=.9\columnwidth]{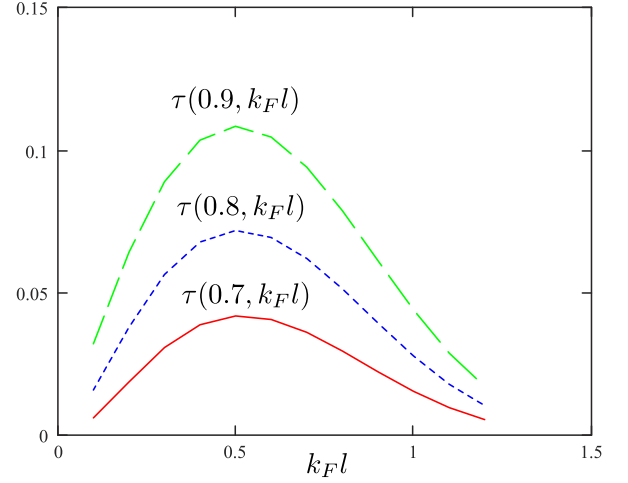}
\end{center}
\caption{The function $\tau(g,x)$ for $g=0.7$ (solid line), $g=0.8$
(short-dashed line), and $g=0.9$ (long-dashed line).}%
\label{Fig_tau}%
\end{figure}

The dependence of the function $\tau(g,k_{F}l)$ on $k_{F}l$ for several values
of $g$ is shown in Fig. \ref{Fig_tau}. We see that the critical temperature decreases very rapidly for
$k_{F}l>1$ due to the fast decay of the scattering amplitude. The optimal
value of $k_{F}l$ is around $0.5$ with the critical temperature reaching
values of the order of $0.1\mu$ for $g\approx0.9$ that corresponds to
$gk_{F}l\approx0.45<1$.

\section{Concluding remarks}

We obtain our results for the superfluid critical temperature using the
mean-field approach. However, as it is well-known, this approach in two
dimensions is only applicable at zero temperature, while at finite temperature
the long-range order is destroyed by phase fluctuations and, therefore, the
mean-field order parameter is zero. In this case, the transition into the
superfluid phase follows the Berezinskii-Kosterlitz-Thouless (BKT) scenario
\cite{Berezinskii,KosterlitzThouless}. In the weak coupling limit,
however, as it was pointed out by Miyake \cite{Miyake}, the difference between
the critical temperature calculated within the mean-field approach $T_{c}$ and
the critical temperature of the BKT transition $T_{\mathrm{BKT}}$ can be
estimated as $T_{c}-T_{\mathrm{BKT}}\sim T_{c}^{2}/\mu$ and, therefore, small
as compared to $T_{c}$. As a result, our mean-field calculations provide a
reliable answer for the critical temperature in the considered weak coupling
regime $a_{d}k_{F}<1$.

Let us now discuss possible physical realizations of the interlayer pairing. In the
experiments with polar molecules, the values of the dipolar length $a_{d}$ are
of the order of $10^{2}\div10^{4}\,\mathrm{nm}$: for a $^{40}\mathrm{K}%
^{87}\mathrm{Rb}$ with currently available $d\approx0.3\,\mathrm{D}$ one has
$a_{d}\approx170\,\mathrm{nm}$ (with $a_{d}\approx600\,\mathrm{nm}$ for the
maximum value $d\approx0.566\,\mathrm{D}$), and for $^{6}\mathrm{Li}%
^{133}\mathrm{Cs}$ with a tunable dipole moment from $d=0.35\,\mathrm{D}$ to
$d=1.3\,\mathrm{D}$ (in an external electric field $\sim1\,\mathrm{kV/cm}$)
the value of $a_{d}$ varies from $a_{d}\approx260\,\mathrm{nm}$ to
$a_{d}\approx3500\,\mathrm{nm}$. For the interlayer separation $l$ of the
order of few hundreds nanometers, the corresponding values of the parameter
$g$ can be both smaller and larger than unity ($g\lesssim10$).

The values of the parameter $k_{F}l$ are also within this range for densities
$n=10^{6}\div10^{9}\,\mathrm{cm}^{-2}$ (for example, one has $k_{F}l=1$ for
$l=500\,\mathrm{nm}$ and $n\approx3\cdot10^{7}\,\mathrm{cm}^{-2}$). Note,
however, that the optimal values of this parameter are around $k_{F}l\sim0.5$
(see Fig. \ref{Fig_tau}) and, hence, the optimum value of the interlayer
separtion is related to the density, which, in turn, should be large enough to
provide a substantial value for the Fermi energy. For$^{40}\mathrm{K}%
^{87}\mathrm{Rb}$ molecules at the density $n\approx4\cdot10^{8}%
\,\mathrm{cm}^{-2}$ in each layer one has $\varepsilon_{F}\approx
100\,\mathrm{nK}$ and $k_{F}=$. Therefore, the interlayer separation $l$
should be relatively small, $l\lesssim150\,\mathrm{nm}$, to meet the optimual
conditions. For $l=150\,\mathrm{nm}$ one then has $g\approx1.1$ (with current
$d\approx0.3\,\mathrm{D}$), $k_{F}l\approx1$, and $T_{c}\approx0.1\varepsilon
_{F}\approx10\,\mathrm{nK}$. Note that stricktly speaking these values of
parameters $g$ and $k_{F}l$ do not correspond to the weak coupling regime
considered in this paper, rather to the intermediate regime of the BCS-BEC
crossover. However, based on the experience with the BEC-BCS crossover in
two-component atomic fermionic mixtures, in which the critical temrature
contines to grow when approaching the crossover region from the BCS side, we
could expect that the above value of the critical temperature provides a good
estimate for the onset of the superfluidity in the intermediate coupling regime.

\section*{Acknowledgements}

We acknowledge fruitful discussions with G. Pupillo, P. Julienne, G.V.
Shlyapnikov, L. Sieberer, and J. Ye.  This work was supported by the Austrian Science Fund FWF (SFB FOQUS),
the EU STREP NAME-QUAM, and the AFOSR MURI.

\appendix

\appendix

\section{Interlayer bound state}\label{appendix_A}



We present in this appendix some details of the calculation of the (intra-layer) bound state properties for small couplings $g\ll1$. Our starting point is Eq. (8), i.e. the equation for the radial wave-function $\chi_{m_z}(\rho)$ of the bound state with binding energy $E_b$. Since for $g\ll1$ one has merely one (shallow) bound state of axial symmetry, in the following we focus on the the axial symmetric case, $m_z=0$, and derive its binding energy and wave-function within a series in $1/g$.\\

On one hand, we see from Eq.~\eqref{Schroedinger} that at sufficiently large distances, cf. $\rho\gg\rho_\star$, we can neglect the interaction potential $V_{2D}(\rho)$ and the wave-function takes the form
\begin{align}\label{eq:A1}
\chi_0(\rho) &\approx {\cal C} K_0(\sqrt{m E_b} \rho/\hbar), 
\end{align}
with ${\cal C}$ a constant and $K_0(z)$ the modified Bessel function of the second kind and 
the distance $\rho_\star\sim(d^2/E_b)^{1/3}=(gl\hbar^2/mE_b)^{1/3}\gg l$ for $g\gg ml^2E_b/\hbar^2$.
Since $\rho_\star\gg\hbar/\sqrt{mE_b}\equiv\rho_\kappa$ for $E_b\ll\hbar^2g/ml^2$, we can expand Eq.~\eqref{eq:A1} for distances $\rho\ll \rho_k$ as
\begin{align}
\chi_0(\rho) &\approx {\cal C} \ln\left(\frac{2\hbar e^{-\gamma}}{\sqrt{mE_b}\rho}\right)\nonumber 
\end{align}
with $\gamma\approx 0.5772$ the Euler constant and in particular 
\begin{align}\label{eq:A2}
\rho\frac{d}{d\rho}\ln[\chi_0(\rho)]\approx
-\left[\ln\left(\frac{2\hbar}{\sqrt{m E_b}\rho}\right)-\gamma\right]^{-1}.
\end{align}

On the other hand, for sufficiently small distances, cf. $\rho\ll\rho_\star$, and weak coupling $g\ll1$, we can neglect the bound state energy, i.e. we assume $E_b\ll\hbar^2g/ml^2$, and expand the wavefunction in powers of $g$ as
\begin{align}
\chi_0(\rho) \approx {\cal N}\left[\chi_0^{(0)}(\rho) + \sum_{n=1}^\infty g^n \chi_0^{(n)}(\rho)\right]\nonumber
\end{align}
with ${\cal N}$ an overall normalization constant. Then Eq.~\eqref{Schroedinger} gives a set of differential equations for the various terms, 
\begin{align}\label{eq:A5}
\frac{1}{\rho}\frac{d}{d \rho}\rho\frac{d}{d \rho}\chi_0^{(n)}(\rho)=\frac{m V_{2D}(\rho)}{\hbar^2 g}\chi_0^{(n-1)}(\rho)
\end{align}
with $\chi_0^{(-1)}(\rho)\equiv0$ and the boundary condition 
 $\chi_0^{(n)}(0)=\delta_{n,0}$.
From Eq.~\eqref{eq:A5} we obtain the zeroth order term is then a constant, i.e. $\chi_0^{(0)}(\rho)=1$, while the higher order terms are obtain by iterated integration of Eq.~\eqref{eq:A5} as
\begin{align}
\chi_0^{(n)}(\rho) =& \int_{0}^{\rho} d\rho_1 \int_{0}^{\rho_1}d\rho_2 \frac{\rho_2l(\rho_2^2-2l^2)}{\rho_1(\rho_2^2+l^2)^{5/2}}\chi_0^{(n-1)}(\rho_2).\nonumber 
\end{align}
The terms up to forth order are (with $z\equiv\sqrt{\rho^2+l^2}/l$)
\begin{subequations}
\begin{align}
\chi_0^{(0)}(\rho)&=1,\nonumber\\
 \chi_0^{(1)}(\rho)&=\frac{1}{z}-1,\nonumber\\
 \chi_0^{(2)}(\rho)&=\frac{3}{8z^2}-\frac{1}{z}+\frac{5}{8}-\frac{1}{4}\ln(z),\nonumber\\
\chi_0^{(3)}(\rho)&=\frac{3}{40z^3}-\frac{3}{8z^2}+\frac{47}{120z}-\frac{11}{120}-\frac{z-1}{4}\ln(z)\nonumber\\
&\quad-\frac{4}{15}\ln\left(\frac{z+1}{2}\right),\nonumber\\
\chi_0^{(4)}(\rho)&=\frac{3}{320 z^4}-\frac{3}{40 z^3}+\frac{11}{128 z^2}+\frac{17}{120 z}-\frac{311}{1920}\nonumber\\
&\quad+\frac{\ln^2(z)}{32}+\left(-\frac{3}{32 z^2}+\frac{1}{4 z}+\frac{257}{960}\right) \ln (z)\nonumber\\
&\quad+\frac{4}{15}\left(\frac{1}{z}+1\right) \ln \left(\frac{2}{z+1}\right)+\frac{{\rm Li}_2\left(1-z^2\right)}{64},\nonumber
\end{align}
\end{subequations}
with ${\rm Li}_n(z)=\sum_{k=1}^\infty z^k/k^n$ the polylogarithm function.

We truncate the latter expansion at order $g^n$ and have
\begin{align}
\rho\frac{d}{d\rho}\ln\chi_0(\rho)\approx \rho\frac{d}{d\rho}\ln\left[\sum_{k=0}^n g^k \chi^{(k)}_0(\rho)\right]\equiv \Lambda_{0}^{(n)}(\rho),\nonumber
\end{align}
which for $\rho\gg l$ (cf. $z\gg1$) gives
\begin{align}
\Lambda_0^{(0)}(\rho)&=0,\nonumber\\
\Lambda_0^{(1)}(\rho)&=-\frac{g}{z^2}\frac{z^2-1}{z+g(1-z)}\approx-\frac{g}{1-g}\frac{l}{\rho},\nonumber\\
\Lambda_0^{(2)}(\rho)&=\left[-\frac{4}{g z^3}+\frac{4}{g z}-\frac{3}{z^4}+\frac{4}{z^3}+\frac{2}{z^2}-\frac{4}{z}+1\right]\nonumber\\
&\quad\times\left[-\frac{4}{g^2}-\frac{4}{g z}+\frac{4}{g}-\frac{3}{2 z^2}+\frac{4}{z}+\ln (z)-\frac{5}{2}\right]^{-1}\nonumber\\
&\approx\left[\ln\left(\frac{\rho}{l}\right)-\left(\frac{4}{g^2}-\frac{4}{g}+\frac{5}{2} \right)\right]^{-1},\nonumber\\
\Lambda_0^{(3)}(\rho)&\approx\left\{\ln\left(\frac{\rho}{l}\right)-\left(1-\frac{g}{15}\right)^{-1}\right.\nonumber\\
&\left.\quad\times\left[\frac{4}{g^2}-\frac{4}{g}+\frac{5}{2} + \frac{32\ln(2)-11}{30}g \right]\right\}^{-1},\nonumber\\
\Lambda_0^{(4)}(\rho)&\approx\left\{\ln\left(\frac{\rho}{l}\right)-\left(1-\frac{g}{15}-\frac{g^2}{240}\right)^{-1}\right.\nonumber\\
&\left.\quad\times\left[\frac{4}{g^2}-\frac{4}{g}+\frac{5}{2} + \frac{32\ln(2)-11}{30}g\right.\right.\nonumber\\
&\quad\quad\left.\left. -\frac{311+5\pi^2-512\ln(2)}{480}g^2\right]\right\}^{-1}, \label{eq:A11}
\end{align}

 Comparing the latter with the asymptotic behavior from Eq.~\eqref{eq:A2}, we see that for $n\geq2$ we can match their leading $\sim 1/\log(\rho)$ behavior via the binding energy and thus obtain (up to order $n$), respectively,
\begin{subequations}\label{eq:A13}
\begin{align}
\ln\left[\frac{ml^2E_b^{(2)}}{4\hbar^2e^{-2\gamma}}\right]&= -\frac{8}{g^2}+\frac{8}{g}-5,\\
\ln\left[\frac{ml^2E_b^{(3)}}{4\hbar^2e^{-2\gamma}}\right]&= -\frac{8/g^2 - 8/g + 5}{1+g/15}-\frac{32\ln(2)-11}{15/g+1}\nonumber\\
&\approx -\frac{8}{g^2}+\frac{128}{15 g}-\frac{1253}{225}+{\cal O}(g),\\
\ln\left[\frac{ml^2E_b^{(4)}}{4\hbar^2e^{-2\gamma}}\right]&=-\frac{8/g^2 - 8/g + 5 - 11g/15-311g^2/240}{1+g/15-g^2/240}\nonumber\\
&\quad-\frac{32g (g + 1)\ln(2)/15 - \pi^2 g^2/48}{1+g/15-g^2/240}\nonumber\\
&\approx-\frac{8}{g^2}+\frac{128}{15 g}-\frac{2521}{450}+{\cal O}(g).
\end{align}
\end{subequations}
We see that the truncation at order $n$ yields the correct expansion for the logarithm of the energy up to ${\cal O}(g^{n-1})$, and thereby obtain from the latter obtain
\begin{align}
E_b \approx \frac{4\hbar^2}{ml^2}\exp\left[-\frac{8}{g^2}+\frac{128}{15 g}-\frac{2521}{450}-2\gamma+{\cal O}(g)\right].\nonumber
\end{align}

\begin{figure}[tb]
\begin{center}
\includegraphics[width=.95\columnwidth]{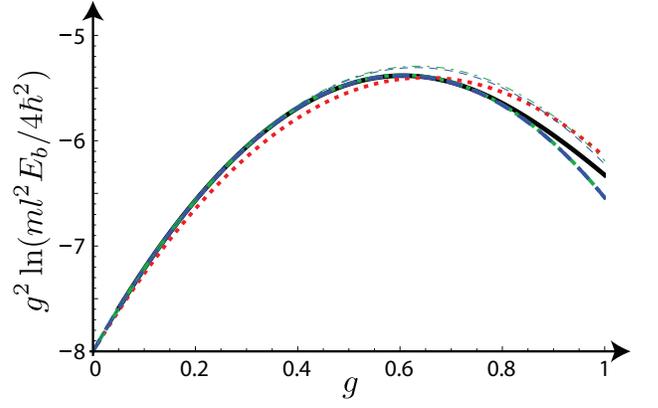}
\end{center}
\caption{\label{Fig_Boundstate1} Binding energy (scaled), $g^2\ln(ml^2 E_b/4\hbar^2)$, as a function of the coupling $g$. Show are the numerical result $E_b^{\rm n}$ (solid line) and the analytical results $E_b^{(n)}$ in $n=2,3,4$-th order perturbation theory (dotted, dot-dashed, dashed lines), as well as the expansion up to ${\cal O}(g)$ for $n=3,4$ (thin dot-dashed, thin dashed lines) from Eq.~\eqref{eq:A13}. Note that the numerical results for $n=3$ and $n=4$ with this scale are practically indistinguishable. See, however, Fig.~\ref{Fig_Boundstate2}.}
\end{figure}

\begin{figure}[tb]
\begin{center}
\includegraphics[width=.95\columnwidth]{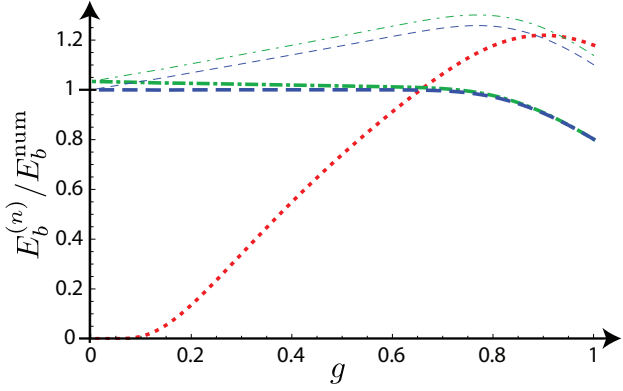}
\end{center}
\caption{\label{Fig_Boundstate2} Deviation of the analytical results from the numerical result for the binding energy, $E^{(n)}_b/E_b^{\rm n}$, as a function of the coupling $g$ for $n=2,3,4$ (dotted, dot-dashed, dashed lines). Also show are the expansion up to ${\cal O}(g)$ for $n=3,4$ (thin dot-dashed, thin dashed lines) from Eq.~\eqref{eq:A13}.}
\end{figure}

Concluding, we compare the analytic results for the binding energy of Eq.~\eqref{eq:A13}, $E_b^{(n)}$, with the result obtain by numerical integration of the Schr\"odinger Eq.~\eqref{Schroedinger}, $E_b^{\rm n}$, which are shown in Fig.~\ref{Fig_Boundstate1} and Fig.~\ref{Fig_Boundstate2}.

\section{Low energy scattering}\label{appendix_B}

In the following we discuss the connection of the bound-state and in particular its binding energy $E_b$ with the scattering amplitude at (very) low energy for weak coupling $g\ll1$. Our starting is the scattering wave function $\psi_{k}^{(+)}({\boldsymbol \rho})$ at energy $E=\hbar k^2/m$ of Section 3. 

On one hand, for $\rho\gg\rho_\star$ it takes the asymptotic form
\begin{align}
\psi_{k}^{(+)}({\boldsymbol \rho})\approx \exp(i{\bf k}\boldsymbol\rho)-\frac{i f_{k}}{4}H_0^{(1)}(k\rho),\nonumber
\end{align}
with  $f_{k}$ the scattering amplitude and $H_0^{(1)}(z)$ is the Hankel function. In the regime $k\rho_\star\ll1$ we can further expand $\psi_{k}^{(+)}({\boldsymbol \rho})$ for $\rho\ll1/k$ as
\begin{align}\label{eq:B1}
\psi_{k}^{(+)}({\boldsymbol \rho})&\approx 1+\frac{f_k}{2\pi}\left[\log\left(\frac{k\rho}{2}\right)+\gamma-\frac{i\pi}{2}\right]
\end{align}

On the other hand, for $\rho\ll\rho_\star$ and weak coupling $g\ll1$ we can neglect the energy, i.e. take $E\rightarrow0$, and expand the wave-function as a power series in $g$ as
\begin{align}
\psi_k^{(+)}\approx {\cal N}\left[\sum_{n=0}^\infty g^n\chi_0^{(n)}(\rho)\right]
\end{align}
where ${\cal N}$ is a normalization constant. The individual terms, $\chi_0^{(n)}(\rho)$, are normalized as $\chi_0^{(n)}(0)=\delta_{n,0}$ and are derived in Appendix~\ref{appendix_A} up to order $n=4$, see Eq.~\eqref{eq:A13}.
For $\rho\gg l$ (but still $\rho\ll\rho_\star$) the wave-function approaches
\begin{align}
\psi_k^{(+)}(\boldsymbol\rho)&\approx {\cal N} \left\{1-g+g^2 \left[\frac{5}{8}-\frac{\log(\rho/l) }{4}\right]\right.\nonumber\\
&\quad\left.+g^3 \left[\frac{4 \log (2)}{15}-\frac{11}{120}-\frac{\log(\rho/l) }{60}\right]\right.\\
&\quad\left.-g^4 \left[\frac{311}{1920}+\frac{\pi ^2}{384}-\frac{4 \log (2)}{15}-\frac{\log(\rho/l) }{960}\right]\right\}, \nonumber
\end{align}
which has the form of Eq.~\eqref{eq:B1} and thus we can match the two asymptotic forms. This is conveniently done in terms of the log-derivative of the wave-function as
\begin{align}
&\rho\frac{\partial}{\partial\rho}\ln[\psi_k^{(+)}(\boldsymbol\rho)]\approx \left[\ln\left(\frac{k\rho}{2}\right)+\gamma+\frac{2\pi}{f_k}-\frac{i\pi}{2}\right]^{-1},\label{eq:B4}\\
&\rho\frac{\partial}{\partial\rho}\ln[\psi_k^{(+)}(\boldsymbol\rho)]\approx \left[\ln\left(\frac{\rho}{l}\right)+\frac{\Lambda(g)}{2}\right]^{-1},\label{eq:B5}
\end{align}
where $\Lambda(g)$ is obtained as in Eq.~\eqref{eq:A11} as
\begin{align}
\Lambda(g)&\approx \frac{-8/g^2+8/g-5+{\cal O}(g)}{1-g/15+g^2/240+{\cal O}(g^3)} \nonumber\\
&\approx -\frac{8}{g^2}+\frac{128}{15g}-\frac{2521}{450}+{\cal O}(g)\approx\ln\left[\frac{ml^2E_b}{4\hbar^2 e^{-2\gamma}}\right],\nonumber
\end{align}
which we recognize as the perturbative expansion for the binding energy. Matching  Eq.~\eqref{eq:B4} and Eq.~\eqref{eq:B5}, we get the scattering amplitude explicitly as
\begin{align}
f_k &\approx \frac{2\pi}{\log(2/kl)-\gamma+\Lambda(g)/2+i\pi/2}= \frac{4\pi}{\log(E_b/E)+i\pi},\nonumber
\end{align}
which recovers the universal low-energy behavior of two-dimensional scattering. Finally, we remark that expanding the latter expression for the scattering amplitude as a power series up to forth order in $g$ we obtain \begin{align}\label{eq:B6}
\frac{f_k}{2\pi}\approx -\frac{g^2}{4} - \frac{4g^3}{15} - \frac{g^4}{16}\left[\log\left(\frac{2i}{kl}\right)-\gamma+\frac{7}{4}\right].
\end{align}

\section{Born series for the s-wave scattering.}

In the following we derive the (s-wave) scattering amplitude within a Born-expansion. We recall the relation of the s-wave scattering amplitude $f_k$ in terms of the vertex function $\Gamma(E,{\bf k},{\bf k}')$ from Eq.~\eqref{ScatteringIntegralEquation},
\begin{align}
f_k &= \int \frac{d\varphi}{2\pi} f_k(\varphi) =\frac{\langle\Gamma(E,{\bf k},{\bf k}^\prime)\rangle_{\varphi,\varphi'}}{\hbar^2/m} =\frac{m}{\hbar^2}\sum_{n=1}^\infty \Gamma_{s}^{(n)}(k),\nonumber
\end{align}
with ${\bf k},{\bf k}'$ on the mass shell, i.e.~$k=k'=\sqrt{mE/\hbar^2}$, the averaging is performed over their azimuthal angles $\varphi$ and $\varphi'$, and the contributions $\Gamma_{s}^{(n)}(k)$ follow from the Born expansion of the vertex-function $\Gamma(E,{\bf k},{\bf k}')$, cf.~Eq.~\eqref{ScatteringForthOrderExpansion}.

For convenience we introduce the s-wave potential
\begin{align}
\tilde V_s(q_1,q_2)&=\langle \tilde V_{\rm 2D}({\bf q}_1-{\bf q}_2)\rangle_{\varphi_1,\varphi_2}=\langle \tilde V_{\rm 2D}({\bf q}_1-{\bf q}_2)\rangle_{\varphi_1}\nonumber\\
&= \int \frac{d\varphi}{2\pi}\tilde V_{\rm 2D}\left(\sqrt{q_1^2+q_2^2-2q_1q_2\cos\varphi}\right)\nonumber\\
&=g\frac{2\pi\hbar^2}{m} l\frac{\partial}{\partial l}\int\frac{d\varphi}{2\pi}e^{-l\sqrt{q_1^2+q_2^2-2q_1q_2\cos\varphi}},
\end{align}
which in particular for $q_1=0$ (or $q_2=0$) gives
\begin{align}
\tilde V_s(q,0)=\tilde V_s(0,q)=\tilde V_{\rm 2D}(q)=-g\frac{2\pi\hbar^2}{m}e^{-ql}ql,
\end{align}
while for equal momenta, $q_1=q_2=q$, gives
\begin{align}
\tilde V_s(q,q)&=
-\frac{2\pi\hbar^2g}{m}2ql \left[{\bf L}_{-1}(2ql)-I_1(2ql)\right],
\end{align}
with ${\bf L}_{n}(z)$ the modified Struve function and $I_n(z)$ modified Bessel function of the first kind.

\begin{figure}[tb]
\begin{center}
\includegraphics[width=.95\columnwidth]{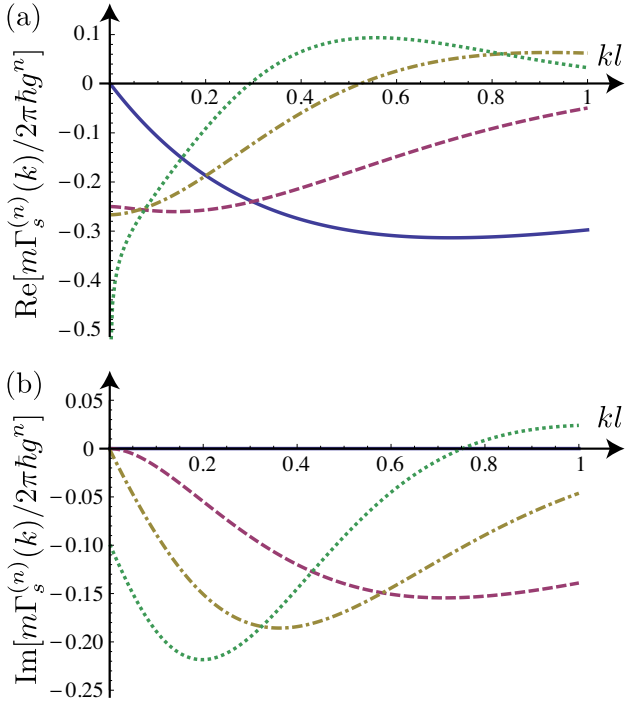}
\end{center}
\caption{\label{fig:C1} Born-series for the s-wave scattering amplitude. Shown are (a) the real and (b) imaginary part of the contributions $\Gamma_s^{(n)}(k)$ of order $n=1,2,3,4$ (solid, dashed, dash-dotted, dotted lines) as a function of momentum $k$.}
\end{figure}

The first order contribution then explicitly gives
\begin{align}
\Gamma_{s}^{(1)}(k)&=\langle V_{2D}({\bf k}-{\bf k}')\rangle_{\varphi_1,\varphi_2} = \tilde V_s(k,k)\nonumber\\
&=-\frac{2\pi\hbar^2g}{m} 2kl \left[{\bf L}_{-1}(2ql)-I_1(2ql)\right],
\end{align}
which vanishes for $k\rightarrow0$, is negative for $k>0$ with its minimum of $\approx-0.3136\times2\pi\hbar^2g/m$ at $k\approx0.7131/l$, see Fig.~\ref{fig:C1} (solid line), and for low-momenta, $k\ll1/l$, gives
\begin{align}
\Gamma_{s}^{(1)}(k)\approx - \frac{2\pi\hbar^2g}{m} \left[\frac{4 kl}{\pi}-2 (kl)^2 +{\cal O}(k^3)\right].
\end{align}

The second order contribution to s-wave scattering is
\begin{align}
\Gamma_s^{(2)}(k)&=\int\frac{d{\bf q}}{(2\pi)^2}\frac{\langle\tilde V_{2D}({\bf k}-{\bf q})\tilde V_{2D}({\bf q}-{\bf k}')\rangle_{\varphi,\varphi'}}{E-\hbar^2 q^2/m+i0^+}\nonumber\\
&=\dashint\frac{qdq}{2\pi}\frac{\tilde V_s(k,q)\tilde V_s(q,k)}{E-\hbar^2 q^2/m}-i\frac{m}{4\hbar^2}\tilde V_s(k,k)^2
\end{align}
The real part, cf. the principal value integral, 
in general has to be evaluated numerically and is shown in Fig.~\ref{fig:C1}(a) (dashed line). We remark that it has two extrema, cf. $\approx-0.2603\times2\pi\hbar^2g^2/m$ at $k\approx0.1364/l$ and $\approx0.0103\times2\pi\hbar^2g^2/m$ at $k\approx2.1137/l$. Moreover for small momenta, $k\ll1/l$, the leading contribution is
\begin{align}
{\rm Re}\left[\Gamma_s^{(2)}(k)\right]
&\approx -\frac{2\pi\hbar^2g^2}{m}l^2\int_0^\infty qdq e^{-2ql}+{\cal O}(k)\nonumber\\
&= -\frac{2\pi\hbar^2g^2}{m}\frac{1}{4}+{\cal O}(k),
\end{align}
which in particular is finite and negative for $k=0$. For the imaginary part we have explicitly
\begin{align}
&{\rm Im}\left[\Gamma_s^{(2)}(k)\right]= -\frac{m}{4\hbar^2} \tilde V_{s}(k,k)^2 
\nonumber\\
&=-\frac{2\pi\hbar^2}{m}  g^2(kl)^22\pi\left[{\bf L}_{-1}(2ql)-I_1(2ql)\right]^2,
\end{align}
with a minimum of $\approx-0.1544\times 2\pi\hbar^2g^2/m$ at $k\approx0.7131/l$, see Fig.~\ref{fig:C1}(b) (dashed line), and for small momenta, $k\ll1/l$, vanishes as
\begin{align}
{\rm Im}\left[\Gamma_s^{(2)}(k)\right]\approx - \frac{16\hbar^2}{m}g^2(kl)^2 +{\cal O}(k^3).
\end{align}

The third order contribution to s-wave scattering is convenitently splits into its real an imaginary part as
\begin{align}
\Gamma_s^{(3)}(k)&=
\int\frac{d{\bf q}_1d{\bf q}_2}{(2\pi)^4}\nonumber\\
&\quad\frac{\langle\tilde V_{\rm2D}({\bf k}-{\bf q}_1) \tilde V_{\rm 2D}({\bf q}_1-{\bf q}_2)\tilde V_{2D}({\bf q}_2-{\bf k}')\rangle_{\varphi,\varphi'}}{(E-\hbar^2 q_1^2/m+i0^+)(E-\hbar^2 q_2^2/m+i0^+)}\nonumber\\
&=\int \frac{q_1 dq_1}{2\pi}\int \frac{q_2 dq_2}{2\pi}\nonumber\\
& \quad\frac{\tilde V_s(k,q_1) \tilde V_s(q_1,q_2) \tilde V_s(q_2,k)}{(E-\hbar^2q_1^2/m+i0^+)(E-\hbar^2q_2^2/m+i0^+)}\nonumber\\
&=\dashint \frac{q_1 dq_1}{2\pi} \dashint \frac{q_2 dq_2}{2\pi}\frac{\tilde V_s(k,q_1) \tilde V_s(q_1,q_2) \tilde V_s(q_2,k)}{(E-\hbar^2q_1^2/m)(E-\hbar^2q_2^2/m)}\nonumber\\
&\quad-\frac{\tilde V_s(k,k)^3}{16\hbar^4/m^2}-i\frac{V_s(k,k)}{2\hbar^2/m}\dashint\frac{qdq}{2\pi}\frac{\tilde V_s(k,q)\tilde V_s(q,k)}{E-\hbar^2 q^2/m},\nonumber
\end{align}
which are shown in Fig.~\ref{fig:C1} (dash-dotted lines). We notice that the real part has its minimum at $k=0$, where
\begin{align}
{\rm Re}\left[\Gamma_s^{(3)}(0)\right]&=\int_0^\infty dq_1\int_0^\infty dq_2\frac{\tilde V_s(0,q_1)\tilde V_s(q_1,q_2)\tilde V_s(q_2,0)}{4\pi^2\hbar^4 q_1 q_2/m^2}\nonumber\\
&=g^2l^2\int_0^\infty dq_1\int_0^\infty dq_2 e^{-q_1l-q_2l}\tilde V_s(q_1,q_2)\nonumber\\
&=2\pi g^2l^2\int_0^\infty \rho d\rho  \frac{ V_{2D}(\rho)}{\rho^2+l^2}
=-\frac{2\pi\hbar^2 g^3}{m} \frac{4}{15},\nonumber
\end{align}
where for the s-wave potential we used the representation 
\[V_s(q_1,q_2)=2\pi\int_0^\infty \rho d\rho V_{2D}(\rho)J_0(q_1\rho)J_0(q_2\rho),\]
and for the convolution of the Bessel function the relation 
\[
\int_0^\infty dq e^{-q l} J_0(q \rho) = \frac{1}{\sqrt{\rho^2+l^2}},\]
and its maximum $\approx0.0636\times2\pi\hbar^2 g^3/m$ at $k\approx0.9205/l$.
While the real part of $\Gamma^{(3)}_s(k)$ is finite and negative for $k=0$,  we notice that its imaginary part vanishes as
\begin{align}
{\rm Im}\left[\Gamma_s^{(3)}(k)\right]&=-\frac{m}{2\hbar^2}\tilde V_s(k,k)~{\rm Re}\left[\Gamma_s^{(2)}(k)\right]\nonumber\\
&\approx -\frac{2\pi\hbar^2g^3}{m}kl +{\cal O}(k^2),
\end{align}
and two extrema, $\approx -0.1857\times 2\pi\hbar^2g^3/m$ at $k\approx0.3623/l$ and $\approx 0.0059\times 2\pi\hbar^2g^3/m$ at $k\approx1.9642/l$.

The forth order contribution to s-wave scattering is conveniently split into its real and imaginary part as
\begin{widetext}
\begin{align}
\Gamma_s^{(4)}(k)&=
\int\frac{d{\bf q}_1d{\bf q}_2d{\bf q}_3}{(2\pi)^6}
\frac{\langle\tilde V_{\rm2D}({\bf k}-{\bf q}_1)V_{\rm 2D}({\bf q}_1-{\bf q}_2)\tilde V_{2D}({\bf q}_2-{\bf q}_3)\tilde V_{2D}({\bf q}_3-{\bf k}')\rangle_{\varphi,\varphi'}}{(E-\hbar^2 q_1^2/m+i0^+)(E-\hbar^2 q_2^2/m+i0^+)(E-\hbar^2 q_3^2/m+i0^+)}\nonumber\\
&=\left(\frac{m}{2\pi\hbar^2}\right)^3\int q_1 dq_1\int q_2 dq_2 \int q_3 dq_3
\frac{V_s(k,q_1)V_s(q_1,q_2)V_s(q_2,q_3)V_s(q_3,k)}{(k^2-q_1^2+i0^+)(k^2-q_2^2+i0^+)(k^2-q_3^2+i0^+)}\nonumber\\
&=\dashint\frac{q_1dq_1}{2\pi}\dashint\frac{q_2dq_2}{2\pi}\dashint\frac{q_3dq_3}{2\pi}
\frac{V_s(k,q_1)V_s(q_1,q_2)V_s(q_2,q_3)V_s(q_3,k)}{(E-\hbar^2q_1^2/m)(E-\hbar^2q_2^2/m)(E-\hbar^2q_3^2/m)}
-3\left[\frac{V_s(k,k)}{4\hbar^2/m}\right]^2 \dashint\frac{qdq}{2\pi}\frac{\tilde V_s(k,q)\tilde V_s(q,k)}{E-\hbar^2 q^2/m}\nonumber\\
&\quad-2i\frac{V_s(k,k)}{4\hbar^2/m}\dashint \frac{q_1 dq_1}{2\pi} \dashint \frac{q_2 dq_2}{2\pi}
\frac{\tilde V_s(k,q_1) \tilde V_s(q_1,q_2) \tilde V_s(q_2,k)}{(E-\hbar^2q_1^2/m)(E-\hbar^2q_2^2/m)}
-i\frac{m}{4\hbar^2}\left[\dashint\frac{qdq}{2\pi}\frac{\tilde V_s(k,q)\tilde V_s(q,k)}{E-\hbar^2 q^2/m}\right]^2+i\frac{V_s(k,k)^4}{(4\hbar^2/m)^3},\nonumber
\end{align}
\end{widetext}
which are shown in Fig.~\ref{fig:C1} (dotted lines). 
We notice that the imaginary part of $\Gamma_s^{(4)}(k)$ in the limit $k\rightarrow0$ is finite,
\begin{align}
{\rm Im}\left[\Gamma_s^{(4)}(0)\right]&=-\frac{m}{4\hbar^2}\left[\dashint\frac{qdq}{2\pi}\frac{\tilde V_s(0,q)\tilde V_s(q,0)}{E-\hbar^2 q^2/m}\right]^2\nonumber\\
&=-\frac{m}{4\hbar^2}\left[\frac{2\pi\hbar^2}{m}\frac{g^2}{4}\right]^2=-\frac{2\pi\hbar^2}{m}\frac{g^4\pi}{32},\nonumber
\end{align}
with two extrema, $\approx-0.2184\times2\pi\hbar^2/m$ at $k\approx1.0532/l$ and $\approx0.0245\times2\pi\hbar^2/m$ at $k\approx1.0532/l$.
The real part of $\Gamma_s^{(4)}(k)$ diverges for $k\rightarrow0$ as (for $k\ll1/l$)
\begin{align}
&{\rm Re}\left[\Gamma_s^{(4)}(k)\right]\approx\left(\frac{m}{2\pi\hbar^2}\right)^3\dashint\frac{q_1dq_1}{k^2-q_1^2}\dashint\frac{q_2dq_2}{k^2-q_2^2}\dashint\frac{q_3dq_3}{k^2-q_3^2}\nonumber\\
& \qquad\tilde V_s(k,q_1) \tilde V_s(q_1,q_2) \tilde V_s(q_2,q_3) \tilde V_s(q_3,k)+{\cal O}(k^2)\nonumber\\
&\approx\left(\frac{m}{2\pi\hbar^2}\right)^3\dashint\frac{q_2dq_2}{k^2-q_2^2}\left[\dashint\frac{dq}{q}\tilde V_s(0,q) \tilde V_s(q,q_2)\right]^2+{\cal O}(k)\nonumber\\
&=\frac{2\pi\hbar^2g^4}{m}\dashint\frac{qdq}{k^2-q^2}\left[\int \rho d\rho l^2\frac{(\rho^2-2l^2)}{(\rho^2+l^2)^3}J_0(q\rho)\right]^2+{\cal O}(k)\nonumber\\
&=\frac{2\pi\hbar^2g^4}{m}\dashint\frac{qdq}{k^2-q^2}\left[\frac{ql}{2}K_1(ql)-\frac{3(ql)^2}{8}K_2(ql)\right]^2+{\cal O}(k)\nonumber\\
&\approx-\frac{2\pi\hbar^2g^4}{m}\frac{1}{16}\left[\ln\left(\frac{2}{kl}\right)+\frac{7}{4}-\gamma\right]+{\cal O}(k),\nonumber
\end{align}
an has two extrema, i.e. $\approx0.0937\times2\pi\hbar^2g^4/m$ at $k\approx0.5574/l$ and $\approx -0.0027\times2\pi\hbar^2g^4/m$ at $k\approx1.8438/l$.

Concluding we remark that summing up the contributions up to forth order in $g$, we obtain for $k\ll 1/l$,
\begin{align}
\Gamma_s(k)&\approx \sum_{n=1}^4 \Gamma_s^{(n)}(k)\approx-\frac{2\pi\hbar^2}{m}\left\{\frac{g^2}{4}+\frac{4g^3}{15}+\right.\nonumber\\
&\left.+\frac{g^4}{16}\left[\ln\left(\frac{2}{kl}\right)+\frac{7}{4}-\gamma+\frac{i\pi}{2}\right]+{\cal O}(k)\right\},\nonumber
\end{align}
which coincides with the expansion of the scattering amplitude in terms of the binding energy up to forth order in $g$ from Eq.~\eqref{eq:B6}.

%
%
%

\section{The $s$-wave on-shell scattering amplitude in the second Born
approximation}

\label{appendix_D}

We present in this appendix some details of the calculations of the $s$-wave
on-shell scattering amplitude in the second Born approximation. Our starting
expression is%
\[
\Gamma^{(2)}(E,\mathbf{k},\mathbf{k}^{\prime})=\int\frac{d\mathbf{q}}%
{(2\pi)^{2}}\frac{\tilde{V}_{2D}(\mathbf{k}-\mathbf{q})\tilde{V}%
_{2D}(\mathbf{q}-\mathbf{k}^{\prime})}{E-q^{2}\hbar^{2}/m+i0},
\]
where $\tilde{V}_{2D}(\mathbf{k}-\mathbf{q})$ is given by Eq.~\eqref{V2DFourier},
$k=k^{\prime}=\sqrt{mE}/\hbar$, and we have to perform averaging over the direction
of $\mathbf{k}$ and $\mathbf{k}^{\prime}$ (azimuthal angles $\varphi$ and
$\varphi^{\prime}$, respectively)%
\[
\Gamma_{s}^{(2)}(k)=\int\frac{d\varphi}{2\pi}\int\frac{d\varphi^{\prime}}%
{2\pi}\Gamma^{(2)}(E,\mathbf{k},\mathbf{k}^{\prime})
\]
in order to obtain the $s$-wave contribution.

The calculation of the imaginary part is simple and can be performed without
the on-shell condition $k=k^{\prime}=q_E$:%
\begin{align*}
&\mathrm{Im}\left[\Gamma^{(2)}(E,\mathbf{k},\mathbf{k}^{\prime})\right]    =-\pi\int
\frac{d\mathbf{q}}{(2\pi)^{2}}\tilde{V}_{2D}(\mathbf{k}-\mathbf{q})\nonumber\\
& \qquad \qquad \qquad\tilde
{V}_{2D}(\mathbf{q}-\mathbf{k}^{\prime})\delta(E-q^{2}\hbar^{2}/m)\\
& \qquad =-\frac{m}{4\hbar^{2}}\int\frac{d\varphi_{\mathbf{q}}}{2\pi}\tilde{V}%
_{2D}(\mathbf{k}-\mathbf{q}_{E})\tilde{V}_{2D}(\mathbf{q}_{E}-\mathbf{k}%
^{\prime}),
\end{align*}
where $q_{E}=\sqrt{mE}/\hbar$. The $s$-wave contribution then is%
\begin{align}
&\mathrm{Im}\left[\Gamma_{s}^{(2)}(E,\mathbf{k},\mathbf{k}^{\prime})\right]\nonumber\\
&=-\frac{m}%
{4\hbar^{2}}\left\langle \tilde{V}_{2D}(\mathbf{k}-\mathbf{q}_{E}%
)\right\rangle _{\varphi}\left\langle \tilde{V}_{2D}(\mathbf{q}_{E}%
-\mathbf{k}^{\prime})\right\rangle _{\varphi^{\prime}},\nonumber
\end{align}
where%
\begin{align}
\left\langle \tilde{V}_{2D}(\mathbf{k}-\mathbf{q}_{E})\right\rangle _{\varphi
}&=\int\frac{d\varphi}{2\pi}\tilde{V}_{2D}(\sqrt{k^{2}+q_{E}^{2}-2kq_{E}%
\cos\varphi}),\nonumber\\
\left\langle \tilde{V}_{2D}(\mathbf{q}_{E}-\mathbf{k}^{\prime})\right\rangle
_{\varphi^{\prime}}&=\int\frac{d\varphi^{\prime}}{2\pi}\tilde{V}_{2D}%
(\sqrt{k^{\prime2}+q_{E}^{2}-2k^{\prime}q_{E}\cos\varphi^{\prime}}).\nonumber
\end{align}
On the mass shell $k=k^{\prime}=q_{E}=\sqrt{mE}/\hbar$ and under the
condition $kl\ll1$, we have%
\[
\left\langle \tilde{V}_{2D}(\mathbf{k}-\mathbf{q}_{E})\right\rangle _{\varphi
}=-\frac{2\pi\hbar^{2}}{m}\frac{4}{\pi}gkl
\]
and, therefore,%
\begin{equation}
\mathrm{Im}\left[\Gamma_{s}^{(2)}(k)\right]=-\frac{2\pi\hbar^{2}}{m}\frac{8}{\pi}%
g^{2}(kl)^{2}. \label{ImGamma2}%
\end{equation}

The calculation of the real part%
\begin{equation}
\mathrm{Re}\left[\Gamma_{s}^{(2)}(k)\right]=\dashint
\frac{d\mathbf{q}}{(2\pi)^{2}}\frac{\langle \tilde{V}_{2D}(\mathbf{k}%
-\mathbf{q}_{E})\rangle _{\varphi}\langle \tilde{V}_{2D}%
(\mathbf{q}-\mathbf{k}^{\prime})\rangle _{\varphi^{\prime}}}{\hbar^2(k^{2}%
-q^{2})/m}, \label{ReGamma2_1}%
\end{equation}
where $\dashint$ denotes the principal value of the integral, is
technically more involved. After introducing the new dimensionless integration
variable $y=q/k$, Eq. (\ref{ReGamma2_1}) reads%
\begin{align}
\mathrm{Re}\left[\Gamma_{s}^{(2)}(k)\right]&=\frac{2\pi\hbar^{2}}{m}g^{2}\dashint_{0}^{\infty}\frac{\varepsilon
^{2}ydy}{1-y^{2}}\left\langle R_{1}%
R_{2}e^{-\varepsilon(R_{1}+R_{2})}\right\rangle _{\varphi_{1}\varphi_{2}}\nonumber
\end{align}
where $\varepsilon=kl\ll1$, $R_{i}=\sqrt{1+y^{2}-2y\cos\varphi_{i}}$,
$\varphi_{1}=\varphi$, and $\varphi_{2}=\varphi^{\prime}$. After integrating
by part we obtain%
\begin{align}
\mathrm{Re}\left[\Gamma_{s}^{(2)}(k)\right]&=\frac{2\pi\hbar^{2}}{m}\frac{g^{2}}%
{2}\varepsilon^{2}\int_{0}^{\infty}dy\ln(\left\vert 1-y^{2}\right\vert
)\nonumber\\
&\qquad\frac{d}{dy}\left\langle R_{1}R_{2}\exp[-\varepsilon(R_{1}+R_{2}%
)]\right\rangle _{\varphi_{1}\varphi_{2}},\nonumber
\end{align}
and the calculation of $\mathrm{Re}[\Gamma_{s}^{(2)}(k)]$ reduces to the
calculation of the integral%
\[
I(\varepsilon)=\varepsilon^{2}\int_{0}^{\infty}dy\ln(\left\vert 1-y^{2}%
\right\vert )\frac{d}{dy}\left\langle R_{1}R_{2}e^{-\varepsilon(R_{1}%
+R_{2})}\right\rangle _{\varphi_{1}\varphi_{2}}%
\]
up to the terms $\sim\varepsilon^{2}\ln\varepsilon$.

It is convenient to split $I(\varepsilon)$ into two parts, i.e $
I(\varepsilon)=I_{1}(\varepsilon)+I_{2}(\varepsilon)$, where%
\begin{align}
I_{1}(\varepsilon)&=\varepsilon^{2}\int_{0}^{\infty}dy\ln(y^{2}-1)\frac{d}%
{dy}\left\langle R_{1}R_{2}e^{-\varepsilon(R_{1}+R_{2})}\right\rangle
_{\varphi_{1}\varphi_{2}},\nonumber\\
I_{2}(\varepsilon)&=\varepsilon^{2}\int_{1}^{1}dy\ln(1-y^{2})\frac{d}%
{dy}\left\langle R_{1}R_{2}e^{-\varepsilon(R_{1}+R_{2})}\right\rangle
_{\varphi_{1}\varphi_{2}}.\nonumber
\end{align}
Within the chosen accuracy, the second integral $I_{2}(\varepsilon)$ we can be
simplified%
\[
I_{1}(\varepsilon)\approx\varepsilon^{2}\int_{0}^{1}dy\ln(1-y^{2})\frac{d}%
{dy}\left\langle R_{1}R_{2}\right\rangle _{\varphi_{1}\varphi_{2}}
\]
because it converges when $\varepsilon\rightarrow0$ in the integrand. The
result of averaging over angles $\varphi_{1}$ and $\varphi_{2}$ can now be
expressed in terms of complete elliptic integrals $\mathrm{E}(k)$ and
$\mathrm{K}(k)$:%
\begin{align}
I_{2}(\varepsilon)&=\varepsilon^{2}\frac{4}{\pi^{2}}\int_{0}^{1}dy\ln
(1-y^{2})\frac{1+y}{y}\mathrm{E}(k)\nonumber\\
& \qquad \qquad\left[  (y+1)\mathrm{E}(k)+(y-1)\mathrm{K}%
(k)\right]  ,
\end{align}
where $k=4y/(1+y)^{2}$. The numerical calculation of the above integral gives%
\begin{equation}
I_{2}(\varepsilon)\approx-0.697\varepsilon^{2}. \label{I2}%
\end{equation}

In order to calculate the integral $I_{2}(\varepsilon)$ we differential first
with respect to $y$:%

\begin{align}
I_{2}(\varepsilon)  &  =\varepsilon^{2}\int_{1}^{\infty}dy\ln(y^{2}%
-1)\left\langle \left\{  \frac{y-\cos\varphi_{1}}{R_{1}/R_2}+%
\frac{y-\cos\varphi_{2}}{R_{2}/R_1}\right.\right.\nonumber\\
& \qquad \quad\left.\left.-\varepsilon\left[  (y-\cos\varphi_{1}%
)R_{2}+(y-\cos\varphi_{2})R_{1}\right]  \right\}\right.\nonumber\\
& \qquad \quad\left.  e^{-\varepsilon
(R_{1}+R_{2})}\right\rangle _{\varphi_{1}\varphi_{2}}=I_{2a}(\varepsilon)+I_{2b}(\varepsilon),
\end{align}
where we split the integral into the two contributions
\begin{align}
I_{2a}(\varepsilon)&=\varepsilon^{2}\int_{1}^{\infty}dy\ln(y^{2}-1)
\left\langle e^{-\varepsilon(R_{1}+R_{2})}\right.\nonumber\\
&\quad\left.\left[  \frac{y-\cos\varphi_{1}}{R_{1}/R_2}+\frac{y-\cos\varphi_{2}%
}{R_{2}/R_1}\right]\right\rangle _{\varphi
_{1}\varphi_{2}}\nonumber,\\
I_{2b}(\varepsilon)&=-\varepsilon^{3}\int_{1}^{\infty}dy\ln(y^{2}%
-1)\left\langle e^{-\varepsilon(R_{1}+R_{2})}\right.\nonumber\\
& \quad\left.\left[  (y-\cos\varphi_{1})R_{2}+R_{1}(y-\cos\varphi
_{2})\right]\right\rangle _{\varphi
_{1}\varphi_{2}}.\nonumber
\end{align}
For $y>1$ we can write%
\begin{align}
&R_{1}+R_{2}\approx \sum_{i=1}^2 \left[y-\cos\varphi_{i}+\frac{\sin
^{2}\varphi_{i}}{2y}\right]+{\cal O}(y^{-2}),\label{expansion1}\\%
&(y-\cos\varphi_{1})R_{2}+R_{1}(y-\cos\varphi_{2})\approx2y^{2}-2y\sum_{i=1}^2\cos
\varphi_i\nonumber\\
&+\frac{\sin^{2}\varphi_{1}+\sin^{2}%
\varphi_{2}+4\cos\varphi_{1}\cos\varphi_{2}}{2}+\mathrm{O}(y^{-1}).
\label{expansion2}%
\end{align}
%
Note that the integral $I_{2b}(\varepsilon)$ is already proportional to
$\varepsilon^{3}$ and, hence, the contribution of finite $y$ ($1\sim
y\ll\varepsilon^{-1}$) will be beyond the necessary accuracy. Therefore, we
should consider only the contribution of large $y$ ($y\sim\varepsilon^{-1}$).
In this case we can replace $\ln(y^{2}-1)$ with $2\ln y$ and use Eqs.
(\ref{expansion1}) and (\ref{expansion2}). In the exponent $\exp
[-\varepsilon(R_{1}+R_{2})]$ we keep only $2\varepsilon y$ from the expansion
for $\varepsilon(R_{1}+R_{2})$ to ensure convergency, while expand in
$\varepsilon(\cos\varphi_{1}+\cos\varphi_{2})$ and $(\varepsilon/2y)(\sin
^{2}\varphi_{1}+\sin^{2}\varphi_{2})$ to the second and the first order,
respectively,%
\begin{align}
&e^{-\varepsilon(R_{1}+R_{2})}\approx e^{-2\varepsilon y}\big[
1+\varepsilon(\cos\varphi_{1}+\cos\varphi_{2})\nonumber\\
&\quad+\frac{\varepsilon^{2}}{2}%
(\cos\varphi_{1}+\cos\varphi_{2})^{2}-\frac{\varepsilon}{2y}(\sin^{2}%
\varphi_{1}+\sin^{2}\varphi_{2})+\ldots\big].\nonumber
\end{align}
It is easy to see that higher order terms in the above expansion, as well as
the omitted terms in Eq. (\ref{expansion2}) result in terms $\sim
\varepsilon^{3}\ln\varepsilon$ or smaller. The integrations over $y$ and the
angles $\varphi_{1}$ and $\varphi_{2}$ are then straightforward and give%
\begin{equation}
I_{2b}(\varepsilon)\approx-\frac{3}{2}+\gamma+\ln(2\varepsilon)+\varepsilon
^{2}\left[  \frac{7}{4}-\frac{1}{2}\gamma-\frac{1}{2}\ln(2\varepsilon)\right]
, \label{I1b}%
\end{equation}
where $\gamma\approx0.5772$ is the Euler constant.

The integral $I_{2a}(\varepsilon)$ can be rewritten in the form%
\begin{align}
I_{2a}(\varepsilon)  &  =I_{2a1}(\varepsilon)+I_{2a2}(\varepsilon)  =\varepsilon^{2}\int_{1}^{\infty}dy\ln(y^{2}-1)e^{-2\varepsilon
y}\nonumber\\
&\qquad\left\langle \left[
\frac{y-\cos\varphi_{1}}{R_{1}}R_{2}+R_{1}\frac{y-\cos\varphi_{2}}{R_{2}%
}\right]  \right\rangle _{\varphi_{1}\varphi_{2}}\nonumber\\
& \quad+\varepsilon^{2}\int_{1}^{\infty}dy\ln(y^{2}-1)\left\langle\left[  e^{-\varepsilon(R_{1}+R_{2})}-e^{-2\varepsilon
y}\right]\right.\nonumber\\
&\qquad\left. \left[
\frac{y-\cos\varphi_{1}}{R_{1}/R_2}+\frac{y-\cos\varphi_{2}}{R_{2}/R_1%
}\right]    \right\rangle _{\varphi_{1}\varphi_{2}}. \label{I1aO}%
\end{align}
In the second integral, $I_{2a2}(\varepsilon)$, we can write [cf.~Eq.~\eqref{expansion1}]%
\begin{align}
&e^{-\varepsilon(R_{1}+R_{2})}-e^{-2\varepsilon y}\approx\varepsilon
 e^{-2\varepsilon y}\left[  \cos\varphi_{1}+\cos\varphi_{2}\right.\nonumber\\
 &\quad\left.+\frac
{\varepsilon}{2}(\cos\varphi_{1}+\cos\varphi_{2})^{2}-\frac{\sin
^{2}\varphi_{1}+\sin^{2}\varphi_{2}}{2y}+\ldots\right],\nonumber\\
&\frac{y-\cos\varphi_{1}}{R_{1}/R_2}+\frac{y-\cos\varphi_{2}}{R_{2}/R_1%
}\approx2y-(\cos\varphi_{1}+\cos\varphi_{2})\nonumber\\
&\quad+\frac{\cos\varphi
_{1}+\cos\varphi_{2}}{2y^{2}}(\cos\varphi_{1}-\cos\varphi_{2})^{2}+\ldots.
\label{expansion3}%
\end{align}
The calculations are now similar to those that lead us to Eq. (\ref{I1b}), and
we obtain%
\begin{align}
&\varepsilon^{2}\int_{1}^{\infty}dy\ln(y^{2}-1)\left\langle \left[
\frac{y-\cos\varphi_{1}}{R_{1}/R_{2}}+\frac{y-\cos\varphi_{2}}{R_{2}/R_{1}%
}\right]\right.\nonumber\\
&\quad  \left.\left[  e^{-\varepsilon(R_{1}+R_{2})}-e^{2\varepsilon
y}\right]  \right\rangle _{\varphi_{1}\varphi_{2}}\approx\frac{\varepsilon
^{2}}{2}[1-3\gamma+3\ln(2\varepsilon)].\nonumber
\end{align}

In order to calculate the first integral, $I_{2a1}(\varepsilon)$, in Eq.~\eqref{I1aO}, we notice that for $y\gg1$%
\begin{align}
&\left\langle \left[  \frac{y-\cos\varphi_{1}}{R_{1}/R_{2}}+\frac
{y-\cos\varphi_{2}}{R_{2}/R_{1}}\right]  \right\rangle _{\varphi_{1}\varphi_{2}%
}\approx2y-\frac{3}{16y^{3}}+\ldots,\nonumber\\
& \qquad \qquad\ln(y^{2}-1)\approx2\ln y-\frac{1}{y^{2}}+\mathrm{O}(y^{-3}),\nonumber
\end{align}
and rewrite $I_{2a1}(\varepsilon)$ as follows%
\begin{align*}
&  \varepsilon^{2}\int_{1}^{\infty}dy\ln(y^{2}-1)e^{-2\varepsilon y}\left\langle \left[
\frac{y-\cos\varphi_{1}}{R_{1}/R_{2}}+\frac{y-\cos\varphi_{2}}{R_{2}/R_{1}%
}\right]  \right\rangle _{\varphi_{1}\varphi_{2}}\\
&  =\varepsilon^{2}\int_{1}^{\infty}dy\left[  2\ln y-\frac{1}{y^{2}}\right]
2ye^{-2\varepsilon y}\nonumber\\
&\quad+\varepsilon^{2}\int_{1}^{\infty}dy\left[  \ln
(y^{2}-1)-2\ln y+\frac{1}{y^{2}}\right]  2ye^{-2\varepsilon y}\\
& \quad+\varepsilon^{2}\int_{1}^{\infty}dy\ln(y^{2}-1)e^{-2\varepsilon y}\nonumber\\
&\qquad\left\langle \left[
\frac{y-\cos\varphi_{1}}{R_{1}/R_2}+\frac{y-\cos\varphi_{2}}{R_{2}/R_{1}%
}\right]  -2y\right\rangle _{\varphi_{1}\varphi_{2}}.
\end{align*}
The second and the third integrals in the right-hand-site of this equation
converge at $y\sim1$ and, therefore, we can replace $e^{-2\varepsilon y}$
with unity. As a result we obtain%
\begin{align}
&  \varepsilon^{2}\int_{1}^{\infty}dy\ln(y^{2}-1)e^{-2\varepsilon y}\left\langle \left[
\frac{y-\cos\varphi_{1}}{R_{1}/R_{2}}+\frac{y-\cos\varphi_{2}}{R_{2}/R_{1}%
}\right]  \right\rangle _{\varphi_{1}\varphi_{2}}\nonumber\\
&  =\varepsilon^{2}\int_{1}^{\infty}dy\left[  2\ln y-\frac{1}{y^{2}}\right]
2y\exp(-2\varepsilon y)\\
&\quad  +\varepsilon^{2}\int_{1}^{\infty}dy\left[  \ln(y^{2}-1)-2\ln y+\frac
{1}{y^{2}}\right]  2y+\varepsilon^{2}\int_{1}^{\infty}dy\nonumber\\
&\qquad\ln(y^{2}-1)\left\langle \left[
\frac{y-\cos\varphi_{1}}{R_{1}/R_{2}}+\frac{y-\cos\varphi_{2}}{R_{2}/R_{1}%
}\right]  -2y\right\rangle _{\varphi_{1}\varphi_{2}}.\nonumber
\end{align}
The integrations over $y$ and angles $\varphi_{1}$ and $\varphi_{2}$ are then
straightforward and we obtain%
\begin{align*}
&  \varepsilon^{2}\int_{1}^{\infty}dy\ln(y^{2}-1)e^{-2\varepsilon y}\left\langle \left[
\frac{y-\cos\varphi_{1}}{R_{1}/R_{2}}+\frac{y-\cos\varphi_{2}}{R_{2}/R_{1}%
}\right]  \right\rangle _{\varphi_{1}\varphi_{2}}\\
&  =1-\gamma-\ln(2\varepsilon)+\varepsilon^{2}[1+2\gamma+2\ln(2\varepsilon)] -\varepsilon^{2}\\
&\quad  +2\varepsilon^{2}\int_{1}^{\infty}dy\ln(1-y^{2})\left(  \frac{2}{\pi^{2}%
}\frac{1+y}{y}\mathrm{E}(k)\left[  (y+1)\mathrm{E}(k)+\right.\right.\nonumber\\
&\qquad\left.\left.+(y-1)\mathrm{K}%
(k)\right]  -y\right) \\
&  \approx1-\gamma-\ln(2\varepsilon)+\varepsilon^{2}\left\{  2\gamma
+2\ln(2\varepsilon)+2\int_{1}^{\infty}dy\ln(1-y^{2})\right.\nonumber\\
&\qquad\left.\left(  \frac{2}{\pi^{2}%
}\frac{1+y}{y}\mathrm{E}(k)\left[  (y+1)\mathrm{E}(k)+(y-1)\mathrm{K}%
(k)\right]  -y\right)  \right\}
\end{align*}
where again $k=4y/(1+y)^{2}$. Numerical evaluation of the remaining integral
gives%
\begin{align}
&2\int_{1}^{\infty}dy\ln(y^{2}-1)\left\{  \frac{2}{\pi^{2}}\frac{1+y}%
{y}\mathrm{E}(k)\left[  (y+1)\mathrm{E}(k)\right.\right.\nonumber\\
&\qquad\left.\left.+(y-1)\mathrm{K}(k)\right]
-y\right\}  \approx0.0384,
\end{align}
and, hence,%
\begin{align}
I_{2a2}(\varepsilon)  &  \approx1-\gamma-\ln(2\varepsilon)+2\varepsilon
^{2}[\gamma+\ln(2\varepsilon)+0.0192]\nonumber\\
&+\frac{1}{2}\varepsilon^{2}%
[1-5\gamma-5\ln(2\varepsilon)]\label{I1a}\\
&  =1-\gamma-\ln(2\varepsilon)+\varepsilon^{2}[2\gamma+2\ln(2\varepsilon
)+0.03834].\nonumber%
\end{align}
Combining together Eqs. (\ref{I2}), (\ref{I1a}), and (\ref{I1b}), we get%
\[
I(\varepsilon)=-\frac{1}{2}+\varepsilon^{2}\left[  3.323+3\ln(2\varepsilon
)\right]  ,
\]
and, as a result,%
\begin{align}
\mathrm{Re}\left[\Gamma_{s}^{(2)}(k) \right]
&  =\frac{2\pi\hbar^{2}}{m}\frac{g^{2}}{2}\left\{  -\frac{1}{2}+(kl)^{2}%
\left[  5.402+3\ln(kl)\right]  \right\}.\nonumber
\end{align}

\end{document}